\documentclass[peerreview,onecolumn,11pt,draftclsnofoot]{IEEEtran}

\usepackage{amsmath}
\usepackage{multirow}
\usepackage{amsfonts}
\usepackage{epsfig}
\usepackage{amssymb}
\usepackage{graphicx}
\usepackage{amssymb,amsmath}
\usepackage{cite}
\usepackage{color,soul}

\begin{document}
\title{Robust Stackelberg game in communication systems}
\author{Saeedeh Parsaeefard, \IEEEmembership{Student Member, IEEE,}
      and Mihaela van der Schaar, \IEEEmembership{Fellow, IEEE,} Ahmad R. Sharafat, \IEEEmembership{Senior Member, IEEE.}
}
\maketitle

\begin{abstract}
This paper studies multi-user communication systems with two groups of users: leaders which possess system information, and followers which have no system information using the formulation of Stackelberg games. In such games, the leaders play and choose their actions based on their information about the system and the followers choose their actions myopically according to their observations of the aggregate impact of other users. However, obtaining the exact value of these parameters is not practical in communication systems. To study the effect of uncertainty and preserve the players' utilities in these conditions, we introduce a robust equilibrium for Stackelberg games. In this framework, the leaders' information and the followers' observations are uncertain parameters, and the leaders and the followers choose their actions by solving the worst-case robust optimizations. We show that the followers' uncertain parameters always increase the leaders' utilities and decrease the followers' utilities. Conversely, the leaders' uncertain information reduces the leaders' utilities and increases the followers' utilities. We illustrate our theoretical results with the numerical results obtained based on the power control games in the interference channels.
\end{abstract}

\section{Introduction}
Stackelberg games provide a general framework to analyze and design hierarchical interactions among rational, self-interested decision makers (players)\cite{basar,NEexistence}. These hierarchical non-cooperative games consist of two groups of players: leaders, which have complete information about the other players, and followers, which have no system information. First, each leader selects its action by solving a bi-level optimization problem which seeks to maximize the leader's utility subject to the followers' actions as estimated by the leader's information. The followers then select their actions according to their observations from the aggregate impact of other users.

Recently, the authors in \cite{newprespetive,knowledgeandlearning,stackelburggame} have formalized the power control problem in interference channels as Stackelberg games. In these papers, the utility of each user is its throughput and its action is its transmission power. The leaders' information contains the direct and interference channel gains of all the users, and the impacts of users on each other are the amount of interference at their receivers. In this system model, the leaders first determine their transmit powers. Next, the followers' receivers measure (observe) the amount of interference and feed back this value to their transmitters. Finally, the followers' transmitters determine their transmission powers based on this estimated value of the interference. The information possessed by the leaders results in an increase in the leaders' throughput as well as the followers' throughput in some cases (see e.g., \cite{newprespetive}).

However, extracting accurate system information either by the leaders or the followers is a key practical issue when implementing Stackelberg games in communication systems because obtain accurate information is costly and difficult. For example in wireless communication systems, fading, channel noise, delayed in feedback channels, and users' mobility cause uncertainty in the players' parameters. Consequently, not knowing the accurate values of these parameters may lead to worse utilities for both the leaders and the followers. Therefore, it is essential to consider these uncertain parameters and apply the robust approach to guarantee the utilities of both leaders and followers.

Robust optimization theory is a branch of applied mathematics to mitigate the impact of uncertain parameters in the solution of optimization problems. In this context, the uncertain parameter is first modeled as a nominal (estimated) value plus an additive error (the uncertain part) \cite{Gershman}. In the next step, the optimization problem with nominal values (referred to the nominal optimization problem) is mapped to another optimization problem (called the robust counterpart) in which the uncertain parameter is a new optimization variable \cite{selecectedrobust,Generalnorm}.

Generally two basic approaches are applied to define this maping\cite{Bental2000,Gershman,selecectedrobust}: Bayesian approaches where the statistics of error is considered and the utility is statistically guaranteed, while the worst-case approaches where the error is assumed to be bounded within a specific region (uncertainty region) and the utility is guaranteed for any realization of error within this region. Both these approaches have been applied in communications, economics, and mathematics to tackle uncertain parameters in Stacklberg games \cite{Stochasticstack,ffoundded,Incentivesdesign,Localrobustness,Bassarincetive}. In this paper, we choose a worst-case approach to guarantee the players' utilities under uncertainty within a uncertainty region. Based on the above terminology, we refer to the Stackelberg game and its equilibrium with nominal parameters as the nominal Stackelberg game (NSG) and the nominal Stackelberg equilibrium (NSE), respectively. Moreover, the Stackelberg game and its equilibrium where the uncertainty exist and robust optimization is applied, we call the robust Stackelberg game (RSG) and the robust Stackelberg equilibrium (RSE), respectively.

The most related work to this paper is \cite{Bassarincetive} which considers one-leader one-follower Stackelberg game and assumes that the leader does not know the exact values of some parameters to estimate the follower's action. By minimizing the second order sensitivity function of the leader's utility with respect to the uncertain parameters, the paper determines the worst-case utility for the leader given its imperfect information. However, to implement the RSG in communication systems, we encounter different challenges: 1) What is the definition of the RSE if the leaders and followers have different uncertain parameters? 2) What is the performance of the system at the RSG compared to that of the NSG? 3) How can we overcome the additional computational complexities involved where the leader needs to solve the bi-level robust optimization problem? 4) How can we generalize the RSG for multiple-leaders and multiple-followers communication scenario?

To answer the above questions, we first distinguish between the players' uncertain parameters. As stated before, in the Stackelberg game, the actions of the leaders and the followers are determined by the leaders' information and the followers' observations about the aggregate impact of other players, respectively. Consequently, we assume that the uncertain parameters include the leaders' information and the followers' observations. Then, we determine two cases for the RSE: in case 1, the followers' observations are noisy while the leaders possess complete and accurate information sets, and in Case 2, the leaders' information sets are uncertain in addition to the followers' noisy observations.

To evaluate the performance of the RSG compared to that of the NSG, we define two criteria: i) the difference between the players' strategies at the RSE and NSE and ii) the difference between the players' utilities at these two equilibria. Our results show that for case 1, the uncertain parameters increase the leaders' utilities and decrease the followers' utilities. The leaders' strategies are increasing functions and the followers' strategies are decreasing functions with respect to the uncertainty region. In contrast, for Case 2, the leaders' uncertain information  decreases the leaders' utilities and increase the followers' utilities. The leaders' strategies are decreasing functions and the followers' strategies are increasing functions with respect to the uncertainty region. For both of these two cases, we derive the conditions (in terms of system parameters and interactions among users) under which the social utility at the RSE increases as compared to that at the NSE.

In this paper, we derive the relation between the players' strategies at the RSE and at the NSE based on system parameters and bounds of the uncertainty region. Based on this, the complexity associated to the leaders solving robust bi-level optimization problems is reduced considerably. We initiate the analysis of the RSE for the one leader-one follower communication scenario, and generalize it to a multiple-leaders and multiple-followers scenario.

The rest of this paper is organized as follows. In Section II, the system model and game formulation are presented, followed by the definition of the uncertain parameters in Section III. In Section IV, two cases for the RSE are introduced and analyzed for the one-leader one-follower scenario. In Section V, we apply our theoretical findings  to power control games and provide numerical results for the RSE. In Section VI, we extend our framework for the RSG to the multiple-leaders and multiple-followers scenario, followed by our conclusions in Section VII.
\section{System setup}
\subsection{Network model}
\begin{figure}[h]
\begin{center}
\includegraphics[height=3.5cm,width=6cm]{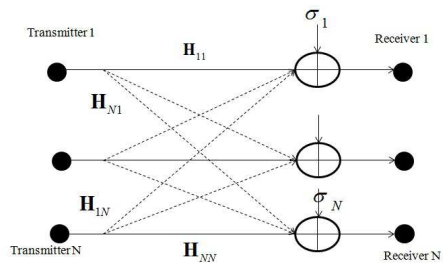}
\end{center}
\caption{Illustrative example: power control games in interference channels.}\label{fig1}
\end{figure}

Consider a set of communication resources divided into $K$ orthogonal dimensions, e.g., frequency bands, time slots, and routes, which are shared between a set of $N$ users. Each user consists of both of a transmitter and receiver. The set of possible positive transmission actions of the $n^{\text{th}}$ user over all the dimensions is given by
\begin{equation*}\label{actionset}
   \mathcal{A}_n=\{ \textbf{a}_n=(a_{n}^{1},\cdots,a_{n}^{K}) | a_{n}^{k} \in [a_{n,k}^{\texttt{min}},a_{n,k}^{\texttt{max}}], \}, \, \forall n \in \mathcal{N}, \,\forall k \in \mathcal{K}
\end{equation*}
where $\mathcal{K}=[1,\cdots,K]$ and $\mathcal{N}=[1,\cdots,N]$. As an example of such a communication system, consider the transmission in the interference channels over $K$ subchannels as depicted in Fig. \ref{fig1}. We denote an achieved utility of user $n$ with $v_n(\textbf{a})$, where $\textbf{a}=[\textbf{a}_n, \textbf{a}_{-n}]$ and $\textbf{a}_{-n}=(\textbf{a}_0,...,\textbf{a}_{n-1},\textbf{a}_{n+1},...,\textbf{a}_{N})$ is a vector of other users' actions except user $n$.

We assume that: A1) the utility function of each user is an increasing, twice differentiable, and concave with respect to $\textbf{a}_n$; A2) the utility function of user $n$ can be expressed as
  \begin{equation}\label{utilityseprale}
    v_n(\textbf{a}_n, \textbf{f}_n(\textbf{a}_{-n}, \textbf{x}_n))=\sum_{k=1}^{K} v_n^k(a_{n}^{k},f_{n}^{k}(\textbf{a}_{-n}, \textbf{x}_n)),
\end{equation}
where $\textbf{f}_n(\textbf{a}_{-n}, \textbf{x}_{n})= [f_{n}^{1}(\textbf{a}_{-n}, \textbf{x}_n),...,f_{n}^{K}(\textbf{a}_{-n}, \textbf{x}_n)]$ is the $1 \times K$ vector of the linear aggregate impact of other users on user $n$ as
\begin{equation}\label{}
        f_{n}^{k}(\textbf{a}_{-n}, \textbf{x}_n)=\sum_{m \in \mathcal{N}, m \neq n} a_{m}^{k}x_{nm}^{k}+y_{n}^{k}
\end{equation}
where $\textbf{x}_n=[\textbf{x}_{n1},\cdots,\textbf{x}_{n(n-1)},\textbf{x}_{n(n+1)}, \cdots,\textbf{x}_{nN}, \textbf{y}_n]$ is the system parameters of user $n$, $\textbf{x}_{nm}$ is the $1 \times K$ vector and $x_{nm}^k$ represents the system parameters between user $m$ and user $n$ in dimension $k$; $\textbf{y}_n=[y_n^1,\cdots, y_n^K]$ where $y_n^k$ denotes the impact of system on user $n$ in dimension $k$. For example, in Fig. \ref{fig1}, $f_n^k(\textbf{a}_{-n}, \textbf{x}_n)$ is the interference of the other users on user $n$ in subchannel $k$  i.e., $f_n^k(\textbf{a}_{-n}, \textbf{x}_n)=\sum_{m\neq n}H_{nm}^k a_{m}^k+\sigma_n^k$, where $H_{nm}^k$ is the channel gain between user $m$ and user $n$ in subchannel $k$, and $a_n^k$ is the transmitted power of user $n$ in subchannel $k$. For this example, the interference channel gains between the users and the noise in each subchannel are the system parameters i.e., $\textbf{x}_{nm}=\textbf{H}_{nm}$  where $\textbf{H}_{nm}=[H^1_{nm},\cdots,H^K_{nm}]$ and $y_{n}^k=\sigma_n^k$.
A3) The utility of user $n$ is a decreasing function of $f_{n}^{k}(\textbf{a}_{-n}, \textbf{x}_n)$; A4) the utility function of each user is twice differentiable over $\textbf{a}_n$, and $\textbf{f}_n(\textbf{a}_{-n}, \textbf{x}_n)$.

Note that A1 is a common assumption for the utility function of all users in the communication systems \cite{CP,ResourceAllocationinWirelessNetworks}. In multiuser communication, A2 and A3 are well-known when the users share the same resources among each other and have a negative impact on each other \cite{mihaelastructure}. A4 indicates the differentiability of utility of each user with respect to $\textbf{a}_n$ and $\textbf{f}_n(\textbf{a}_{-n}, \textbf{x}_n)$. Therefore, all of the above assumptions are practical conditions for communication systems.

This setup includes different game theoretic formulations of communication systems such as additively coupled sum constrained games \cite{mihaelastructure}, which can model many communication systems for example, i) cellular transmission within a given cell, ii) adhoc wireless networks transmission. In this paper, we choose power control games as an illustrative example, where the throughput of each user in the system is its utility i.e., $v_n(\textbf{a}_n, \textbf{f}_n(\textbf{a}_{-n}, \textbf{x}_n))=\sum_{k=1}^{K} \log (1+\frac{H_{nn}^k a_{n}^k}{f^{k}_n(\textbf{a}_{-n}, \textbf{x}_n)})$.

The information set obtained by user $n$ is denoted by $\boldsymbol{\mathcal{I}}_n$, and it may be empty or contain all the private information of other users in the system such as their utilities, and their system parameters. Also the information set of each user may be different from the others. The users with empty and non-empty information sets are referred to as uninformed or informed users, respectively.

Clearly, $\frac{\partial v^k_{n}(a^k_{n}, f^k_{n}(\textbf{a}_{-n}, \textbf{x}_n))}{\partial a_{n}^k}$ is the rate of change of utility of user $n$ corresponding to its action which has a positive value based on A1. A larger value of $\frac{\partial v^k_{n}(a^k_{n}, f^k_{n}(\textbf{a}_{-n}, \textbf{x}_n))}{\partial a_{n}^k}$ means a larger rate of increase of the $n^{\text{th}}$ user's utility with respect to its action. For example, in power control games, we have
\begin{equation}\label{directeffect}
    \frac{\partial v^k_{n}(a^k_{n}, f^k_{n}(\textbf{a}_{-n}, \textbf{x}_n))}{\partial a_{n}^k}= \frac{H_{nn}^{k}}{f_{n}^{k}+H_{nn}^{k}a_{n}^{k}}, \forall n \in \mathcal{N}, \quad \forall k \in \mathcal{K}.
\end{equation}
When the user has a larger direct channel gain i.e., $H_{nn}^{k}\gg  1$  or it has a small interference i.e., $f_{n}^{k} \ll 1$, this value is large. This means that a small change in the user's action causes a large change in the user's utility. We consider the column gradient vector of $v_n$ for user $n$, represented by $\textbf{J}^n_{\textbf{a}_n}=\nabla_{\textbf{a}_n} v_{n}(\textbf{a}_{n}, \textbf{f}_{n}(\textbf{a}_{-n}, \textbf{x}_n))$ as an internal rate of user $n$. Also, let define
\begin{equation}\label{NEGATIVErate}
   \textbf{C}_{nm}=\textbf{X}_{nm} \textbf{J}^n_{\textbf{f}_n},
\end{equation}
where $\textbf{J}^n_{\textbf{f}_n}=\nabla_{\textbf{f}_n} v_{n}(\textbf{a}_{n}, \textbf{f}_{n}(\textbf{a}_{-n}, \textbf{x}_n))$ and $\textbf{X}_{nm}\triangleq \text{diag} \{(x_{nm}^k)_{k=1}^{K}\}$. For example, in power control games,
\begin{equation}\label{directeffect}
    C_{nm}^{k}= -\frac{H_{nn}^{k}H_{nm}^{k}a_{n}^{k}}{f_{n}^{k}(f_{n}^{k}+H_{nn}^{k}a_{n}^{k})}, \forall n \in \mathcal{N}, \quad \forall k \in \mathcal{K},
\end{equation}
where $C_{nm}^k$ is the $k^{\text{th}}$ element of $\textbf{C}_{nm}$. Clearly, larger values of $H_{nm}^{k}$ lead to larger values of $C_{nm}^k$ or larger impact of user $m$ on user $n$. This example shows that $\textbf{C}_{nm}$ denotes the rate of decrease in utility of user $n$ which is linearly related to a corresponding increase in the action of user $m$. We call $\textbf{C}_{nm}$ as the negative impact of user $m$ on user $n$. In the following sections, we use $\textbf{C}_{nm}$ and $\textbf{J}^n_{\textbf{a}_n}$ to study the effect of robustness on the social utility at the RSE compared to the NSE.
\subsection{Game formulation}
Now we model the interaction between informed and uninformed users in communication systems as a Stackelberg game. Assume that the leaders and followers of the Stackelberg game are denoted by $\mathcal{N}_l=\{1,\cdots N_l\}$ and $\mathcal{N}_f=\{1,\cdots N_f\}$, respectively and, $\mathcal{N}=\mathcal{N}_l \cup \mathcal{N}_f$ is the set of all players in the game, where
\begin{eqnarray}
\boldsymbol{\mathcal{I}}_n=&& \{(\mathcal{A}_m, v_{m}, \textbf{X}_{mn},\textbf{X}_{mm}, \textbf{X}_{nm})_{m \neq n, \forall m \in \mathcal{N}}\}  \, \text{if} \, n  \in \mathcal{N}_l  \nonumber\\
  \boldsymbol{\mathcal{I}}_n = && \quad \emptyset \, \quad  \text{if} \quad  n  \in \mathcal{N}_f \nonumber
\end{eqnarray}
where $\textbf{X}_{mn}\triangleq \text{diag} \{(x_{mn}^k)_{k=1}^{K}\}$ and $\textbf{X}_{mm}\triangleq \text{diag} \{(x_{mm}^k)_{k=1}^{K}\}$. In the Stackelberg game, the leaders first play their strategy. Then, the receiver of follower $n$ measures the value of $\textbf{f}_n(\textbf{a}_{-n}, \textbf{x}_n)$ and sends it to its corresponding transmitter to decide its action. Since the value of $\textbf{f}_n(\textbf{a}_{-n}, \textbf{x}_n)$ can be considered as the observation of follower $n$ from the aggregate impact of other users, we refer to it as an observation of user $n$. Therefore, the followers' optimization problems can be formalized,
\begin{equation}\label{utilityseprale}
   \max_{\textbf{a}_n \in \mathcal{A}_n} v_n(\textbf{a}_n, \textbf{f}_n(\textbf{a}_{-n}, \textbf{x}_n)), \quad \, \forall n \in \mathcal{N}_f
\end{equation}
The solution to (\ref{utilityseprale}) for user $n$ represents its best response, denoted by $\textbf{a}^*_{n}(\textbf{a}_{-n})$. Since the followers are non-cooperative players, the equilibrium point of the game emerging among them is $\textbf{a}^{*NE}=(\textbf{a}_{1}^{*},\cdots,\textbf{a}^{*}_{N_f})$, which satisfies
\begin{equation}\label{ne}
    v_n(\textbf{a}^*_{n},\textbf{f}_n(\textbf{a}^*_{-n},\textbf{x}_n))\geq
v_n(\textbf{a}_{n},\textbf{f}_n(\textbf{a}^*_{-n},\textbf{x}_n)) \quad \quad \textbf{a}^*_{n} \in \mathcal{A}_n, 
\end{equation}
where $\textbf{a}^*_{-n}=[\textbf{a}^*_{0},...,\textbf{a}^*_{n-1},\textbf{a}^*_{n+1},...,\textbf{a}^*_{N}]$ for all $n \in \mathcal{N}$. If $\boldsymbol{\mathcal{I}}_n = \emptyset$ for all players, the game reduces to a strategic non-cooperative game.

To define the equilibrium of the Stackelberg game, we consider that there is only one leader in the Stackelberg game which has index $0$. The Stackelberg game equilibrium prescribes an optimal strategy for the leader if the followers play their NE. For example, in a one-leader one-follower Stackelberg game, when player $1$ is a follower, an action $\textbf{a}_0^*$ is the Stackelberg Equilibrium strategy for the player $0$ if $\textbf{a}_0 \in \mathcal{A}_0$ we have
\begin{equation}\label{NasHST}
   v_0(\textbf{a}^{*NSE}_0, \textbf{f}_0(\textbf{a}^{*}_1(\textbf{a}^{*NSE}_0), \textbf{x}_n))\geq v_0(\textbf{a}_0, \textbf{f}_0(\textbf{a}^*_1(\textbf{a}_0), \textbf{x}_n)).
\end{equation}
In this case, the leader's optimization problem changes to the following bi-level optimization
\begin{eqnarray}\label{nominalSTE}
    \max_{\textbf{a}_{0} \in \mathcal{A}_{0}} v_{0}(\textbf{a}_{0},
    \textbf{f}_0(\textbf{a}_{1},\textbf{x}_n)), \quad \forall n \in \mathcal{N} \\
    \text{subject to:} \quad  \max_{\textbf{a}_{1} \in \mathcal{A}_{1}} v_1(\textbf{a}_{1}, \textbf{f}_1(\textbf{a}_{0},\textbf{x}_n)). \nonumber
\end{eqnarray}
For multi-follower scenario, let $\textbf{a}^*_{-0}(\textbf{a}_0)=[\textbf{a}^*_{1},\cdots,\textbf{a}^*_{N-1}]$ be the NE strategy of the followers if player $0$ plays $\textbf{a}_0$. The strategy profile $(\textbf{a}^{*NSE}_0, \textbf{a}^{*NSE}_{-0}(\textbf{a}^{*NSE}_0))$ is the Stackelberg game equilibrium iff
\begin{eqnarray}\label{Multiple}
   v_0(\textbf{a}^{*NSE}_0, \textbf{f}_0(\textbf{a}^{*NSE}_{-0}(\textbf{a}^{*NSE}_0),\textbf{x}_0)) \geq  v_0(\textbf{a}_0, \textbf{f}(\textbf{a}^*_{-0}(\textbf{a}_0),\textbf{x}_0)) , \forall \textbf{a}_0 \in \mathcal{A}_0, \nonumber
\end{eqnarray}
The achieved utility of player $n$ at the NSE and the social utility of the game are denoted by $\omega^{NSE}_{n}$ and $\omega^{NSE}=\sum_{n \in \mathcal{N}}\omega^{NSE}_{n}$, respectively.

Note that if the followers' game has multiple Nash equilibria, the definition of the Stackelberg equilibrium is more complicated as described in \cite{ref1,NEexistence,Convergenceofiterativewaterfillingalgorithm,bilevel}. In this paper we restrict our study to the Stackelberg game with a unique NE in the followers' game. The uniqueness condition for this game is provided in Section VI. In the following, we define the uncertain parameters in the system and based on them, we introduce different types of RSE and the robust counterpart optimization problems for both leaders and followers.
\section{Uncertain parameters}
As stated before, both the followers' observations and the leaders' information sets are uncertain parameters in the considered communication scenario. In the following subsections, we define the followers' uncertain observations and the leaders' uncertain information set which we  noisy observations and incomplete information sets, respectively.

\subsection{Noisy observation} Consider the uncertain value of $\textbf{f}_n(\textbf{a}_{-n},\textbf{x}_n)$ as noisy observation of user $n$ of the impact of the other users, which is modeled as the summation of a deterministic nominal value and an error \cite{saeedeh5}, i.e.,
\begin{equation}\label{modelofucneratinty}
    \widetilde{\textbf{f}}_{n}(\textbf{a}_{-n},\textbf{x}_n)=\textbf{f}_{n}(\textbf{a}_{-n},\textbf{x}_n)+\widehat{\textbf{f}}_{n}(\textbf{a}_{-n},\textbf{x}_n),
\end{equation}
where $\widetilde{\textbf{f}}_{n}(\textbf{a}_{-n},\textbf{x}_n)=[\widetilde{f}^1_{n}(\textbf{a}_{-n},\textbf{x}_n), \cdots \widetilde{f}^K_{n}(\textbf{a}_{-n},\textbf{x}_n)]$, $\textbf{f}_{n}(\textbf{a}_{-n},\textbf{x}_n)=[f^1_{n}(\textbf{a}_{-n},\textbf{x}_n), \cdots f^K_{n}(\textbf{a}_{-n},\textbf{x}_n)]$, and $\widehat{\textbf{f}}_{n}(\textbf{a}_{-n},\textbf{x}_n)=[\widehat{f}^1_{n}(\textbf{a}_{-n},\textbf{x}_n), \cdots \widehat{f}^K_{n}(\textbf{a}_{-n},\textbf{x}_n)]$ are the actual, nominal and the error of the observation of user $n$, respectively. In the worst-case robust optimization theory, the error of noisy observation is assumed to be bounded in a closed region named an uncertainty region \cite{saeedeh5,Robustgame}:
\begin{eqnarray}\label{III-1}
\Re_{n}(\textbf{a}_{-n})=\{ \widetilde{\textbf{f}}_{n}(\textbf{a}_{-n},\textbf{x}_n) \in \Re_{n}(\textbf{a}_{-n}) |  \| \widehat{\textbf{f}}_{n}(\textbf{a}_{-n},\textbf{x}_n) \|_2 \leq \varepsilon_{n} \} \quad , \forall n  \in \mathcal{N}_f
 \end{eqnarray}
where $\varepsilon_n$ is the bound of uncertainty region, and $\|.\|_2$ is the ellipsoid norm. The noisy observation is considered as a new optimization parameter in the utility of each users \cite{saeedeh5}. The new utility function of user $n$, considering the uncertainty, is  $u_n(\textbf{a}_n, \widetilde{\textbf{f}}_{n})$, which satisfies
 \begin{equation}\label{uncertainty}
  u_n(\textbf{a}_n, \widetilde{\textbf{f}}_{n} (\textbf{a}_{-n},\textbf{x}_n))|_{\varepsilon_n=0}=v_n(\textbf{a}_n, \textbf{f}_{n}(\textbf{a}_{-n},\textbf{x}_n)).
 \end{equation}
Consequently, the followers' optimization problem changes to
\begin{equation}\label{utilityrobustfollower}
    \max_{\textbf{a}_n \in \mathcal{A}_n} \min_{ \widetilde{\textbf{f}}_{n}(\textbf{a}_{-n},\textbf{x}_n) \in \Re_{n}(\textbf{a}_{-n})} u_{n}(\textbf{a}_n, \widetilde{\textbf{f}}_{n}(\textbf{a}_{-n},\textbf{x}_n)), \quad \forall n \in \mathcal{N}_f
\end{equation}
The robust Nash Equilibrium (RNE) of this game by assuming the $\boldsymbol{\mathcal{I}}_n = \emptyset$ for all users, is defined \cite{Robustgame,saeedeh5} as $\widetilde{\textbf{a}}^{*}=(\widetilde{\textbf{a}}_{0}^{*},\cdots,\widetilde{\textbf{a}}^{*}_{N-1})$
iff,
\begin{eqnarray}\label{RNEdefinition}
 \min_{ \widetilde{\textbf{f}}_{n}(\textbf{a}^*_{-n},\textbf{x}_n) \in \Re_{n}(\textbf{a}_{-n})} u_{n}(\textbf{a}^*_n, \widetilde{\textbf{f}}_{n}(\textbf{a}^*_{-n},\textbf{x}_n)) \geq \min_{ \widetilde{\textbf{f}}_{n}(\textbf{a}^*_{-n},\textbf{x}_n) \in \Re_{n}(\textbf{a}_{-n})} u_{n}(\textbf{a}_n, \widetilde{\textbf{f}}_{n}(\textbf{a}^*_{-n},\textbf{x}_n) ) \,\,\,\, \forall \textbf{a}_n \in \mathcal{A}_n.
\end{eqnarray}
\subsection{Incomplete information set} We refer to the leaders' information set with uncertain parameters as an incomplete information set where $\textbf{X}_{n_f n_l}$ is the uncertain parameter. Note that obtaining this parameter is more challenging compared to obtaing others parameters in communication systems. For example, in the power control games, the follower's transmitter sends the pilot signals to its receiver to estimate its direct channel gains. The leader can extract $\textbf{H}_{n_fn_f}$ by listening to the follower's feedback channel. Also the leader can estimate  $\textbf{H}_{n_l n_f}$ by listening to this pilot signal. Since there is no pilot signal from the follower's receiver, the leader cannot estimate  $\textbf{H}_{n_f n_l}$. Following the worst-case approach, the uncertain information is considered as the summation of its nominal value and the uncertain part which is bounded in the uncertainty region as
\begin{eqnarray}\label{unceratintyregioninformation}
    \mathcal{R}_{\textbf{X}_{n_f n_l}}=\{ \widetilde{\textbf{X}}_{n_f n_l} \, | \| \widehat{\textbf{X}}_{n_f n_l} \|_2= \|\widetilde{\textbf{X}}_{n_f n_l}- \textbf{X}_{n_f n_l}\|_2 \leq \delta_{n_f n_l} \}, \, n_f  \in \mathcal{N}_f,\, n_l \in \mathcal{N}_l.
\end{eqnarray}
In contrast to the incomplete information set, we refer to the leader's information set without uncertainty as the complete information set.
\section{Robust Stackelberg Equilibrium}
Using the above definitions of uncertain parameters, we can defined different forms of RSG. In this paper, we focus on two common scenarios:
\begin{description}
  \item[Case 1]:  the leaders possess complete information sets, while the followers possess noisy observations;
  \item[Case 2]:  the leaders possess incomplete information sets and the followers possess noisy observations.
\end{description}
As an initial point to analyze the RSE, we first study the one leader-one follower scenario. Then the extension of RSE for multi-followers and multiple-leaders is provided in Section VI.
\subsection{Analysis of RSE for case 1}
For this case, the leader's information set is denote by $\boldsymbol{\mathcal{I}}^{RSE1}_0$ as,
\begin{eqnarray}\label{informationsetRSE}
   \boldsymbol{\mathcal{I}}^{RSE1}_0=\{(\mathcal{A}_m, v_{m}, \textbf{X}_{mn},\textbf{X}_{mm}, \textbf{X}_{nm}, \Re_{m}(\textbf{a}_{-m}))_{m \neq n}, \forall m \in \mathcal{N}\}, \
\end{eqnarray}
Also the leader knows that the follower's optimization problem is the same as (\ref{utilityrobustfollower}). Therefore, the leader's bi-level optimization problem is
\begin{eqnarray}\label{case21}
 &  \max_{\textbf{a}_0 \in \mathcal{A}_0} & v_0(\textbf{a}_0,\textbf{f}_0(\textbf{a}_1, \textbf{x}_{01})) \\
    \text{subject to:}   &\max_{\textbf{a}_n \in \mathcal{A}_n} &\min_{ \widetilde{\textbf{f}}_{1} \in \Re_{1}(\textbf{a}_{-1})} u_{1}(\textbf{a}_1, \widetilde{\textbf{f}}_{1}(\textbf{a}_0, \textbf{x}_{10}) ). \nonumber
\end{eqnarray}
If $\widetilde{\textbf{a}}^*_1(\textbf{a}_0)$ is the best response of (\ref{utilityrobustfollower}) to the leader's action, the RSE for case 1 is $\textbf{a}^{*RSE1}_0$, defined as
\begin{equation}\label{NasHST}
    v_0(\textbf{a}^{*RSE1}_0,\widetilde{\textbf{f}}_{0}(\widetilde{\textbf{a}}^*_1(\textbf{a}^{*RSE1}_0),\textbf{x}_0))\geq v_0(\textbf{a}_0,  \widetilde{\textbf{f}}_{0}(\widetilde{\textbf{a}}^*_1(\textbf{a}_0),\textbf{x}_0)) ,
\end{equation}
\textbf{Remark 1:} The RSE of case 1 exists since: 1) (\ref{utilityrobustfollower}) is a concave with respect to $\textbf{a}_1(\textbf{a}_0)$ for any fixed action of the leader, and a decreasing function with respect to $\textbf{f}_1(\textbf{a}_0,\textbf{x}_1)$, 2) $\mathcal{A}_1$ and $\Re_{1}(\textbf{a}_{-1})$ are two convex bounded and disjoint sets. Consequently, there always exists a saddle point of (\ref{utilityrobustfollower}) \cite{wosrtcasepalomar}, which is a solution to (\ref{utilityrobustfollower}).

Now we characterize and compute the RSE, which is the most difficult part due to the computational complexity of (\ref{case21}). In the following, for notational convenience, we omit the arguments of $\textbf{f}_1(\textbf{a}_0,\textbf{x}_1)$ and $\textbf{f}_0(\textbf{a}_0,\textbf{x}_0)$.

\textbf{Lemma 1}: The uncertain observation for the follower's optimization problem is equal to
\begin{equation}\label{followerstartegyspascecase1}
  \widetilde{\textbf{f}}^{*}_{1}=\textbf{f}_1- \varepsilon_1 \boldsymbol{\vartheta}_{1}
\end{equation}
where $\widetilde{\textbf{f}}_1^*=[\widetilde{f}^{1*}_{1},\cdots, \widetilde{f}^{K*}_{1}]$, $\boldsymbol{\vartheta}_{1}=[\vartheta^1_{1}, \cdots, \vartheta^K_{1}]$, and $\vartheta^k_{1}$ is defined as
\begin{equation}\label{vertana}
    \vartheta^k_{1} =\frac{\frac{\partial u^k_1(\textbf{a}_1, \widetilde{\textbf{f}}_1)}{\partial f_1^k}}{\sqrt{\sum_{k=1}^{K} (\frac{\partial u^k_1(\textbf{a}_1, \widetilde{\textbf{f}}_1)}{\partial f_1^k})^2}},
\end{equation}
\begin{proof}
See Appendix A.
\end{proof}

Using (\ref{followerstartegyspascecase1}) in problem (\ref{case21}) removes the uncertainty region from the leader's bi-level optimization problem which is simplified as
\begin{eqnarray}\label{nominalSTEcase1}
    \max_{\textbf{a}_{0} \in \mathcal{A}_{0}} v_{0}(\textbf{a}_{0},
   \textbf{f}_0), \quad \forall n \in \mathcal{N} \\
    \text{subject to:} \quad  \max_{\textbf{a}_{1} \in \mathcal{A}_{1}} v_1(\textbf{a}_{1}, \widetilde{\textbf{f}}_1^*), \nonumber
\end{eqnarray}
By this reformulation, we can derive the difference between strategy of the leader and the follower in RSE for case 1 and NSE.

\textbf{Proposition 1:} For the case 1 of RSG, the leader's action is an increasing function and the follower's action is a decreasing function with respect to $\varepsilon_1$, and they can be calculated as
\begin{eqnarray} \label{RSE1follower}
  \textbf{a}^{*RSE1}_1&=& \textbf{a}^{*NSE}_1 - \varepsilon_1 \times ((\textbf{J}^{1}_{\textbf{a}_1 \textbf{a}_1})^{-1} \textbf{J}^{1}_{\textbf{a}_1 \textbf{f}_1} \times \vartheta_{1}^{{\scriptsize{\textnormal{T}}}})^{{\scriptsize{\textnormal{T}}}}, \\ \label{RSE1leader}
  \textbf{a}^{*RSE1}_0 &=& \textbf{a}^{*NSE}_0+ \varepsilon_1 \times ((\textbf{J}^{0}_{\textbf{a}_0,\textbf{a}_0})^{-1}\textbf{J}^{0}_{\textbf{a}_0 ,\textbf{f}_0}\boldsymbol{X}_{01} (\textbf{J}^{1}_{\textbf{a}_1 \textbf{a}_1})^{-1} \textbf{J}^{1}_{\textbf{a}_1 \textbf{f}_1} \times \vartheta_{1}^{{\scriptsize{\textnormal{T}}}})^{{\scriptsize{\textnormal{T}}}}.
\end{eqnarray}
where $ \textbf{J}^{n}_{\textbf{f}_n,\textbf{a}_n}= \nabla_{\textbf{f}_{n}\textbf{a}_n } v_{n}(\textbf{a}_{n}, \textbf{f}_{n})$ and $\textbf{J}^{n}_{\textbf{a}_n,\textbf{a}_n}= \nabla^{2}_{\textbf{a}_n} v_{n}(\textbf{a}_{n}, \textbf{f}_{n})$.
\begin{proof}
See Appendix B.
\end{proof}
From Proposition 1, the solution to the robust problem (\ref{case21}) can be calculated based on the NSE and the bound of uncertainty region in (\ref{RSE1follower}) and (\ref{RSE1leader}). Therefore, the computational complexity to solve (\ref{case21}) is reduced. From (\ref{nominalSTEcase1}) and Proposition 1, the variation of utilities of the leader and the follower are derived.

\textbf{Proposition 2:} For the case 1 of RSG, 1) for any realization of the follower's noisy observation, we have,
\begin{eqnarray}
\omega_{0}^{*NSE} \leq \omega_0^{*RSE1}  \quad,
\omega_1^{*RSE1} \leq \omega_{1}^{*NSE} \nonumber
\end{eqnarray}
where $\omega^{*RSE1}_{n}$ is the achieved utility of player $n$ in RSE for case 1. 2) The social utility of game increases compared to NSG, i.e., $\tilde{\omega}^{*RSE1}> \omega^{*NSE}$, if we have
\begin{eqnarray}\label{lemma15-a}
 \text{C1}:  \, |\textbf{C}_{10}| < |\textbf{J}^{0}_{\textbf{a}_0}| \qquad , \text{C2}: \, |\textbf{J}^{1}_{\textbf{a}_1}|< |\textbf{C}_{01}|, \nonumber
\end{eqnarray}
where $|\textbf{q}|$ is the absolute value of the elements of $\textbf{q}$.
\begin{proof}
See Appendix C.
\end{proof}
From Proposition 2, uncertainty in the follower's observation increases the leader's utility. In contrast, the follower achieves a smaller utility in RSE for case 1 compared to NSE. Interestingly, the social utility at the RSE increases compared to NSE if $\text{C1}$ and $\text{C2}$ hold. $\text{C1}$ and $\text{C2}$ can be interpret as follows: the negative impact of the leader to the follower is less than the leader's direct rate and the follower's direct rate is less than its negative impact on the leader. Correspondingly, the increase in the leader's utility is larger than the decrease of the follower's utility where the leader's strategy increases and the follower's strategy decreases and, as a result, the social utility increases.
\subsection{Analysis of RSE for case 2}
For case 2, we denote the leader's incomplete information set as  $\widetilde{\boldsymbol{\mathcal{I}}}^{RSE2}_0$ which is equivalent to
\begin{eqnarray}\label{informationsetRSE2}
\widetilde{\boldsymbol{\mathcal{I}}}^{RSE2}_0=\{(\mathcal{A}_m, v_{m}, \widetilde{\textbf{X}}_{mn}, \textbf{X}_{mm}, \textbf{X}_{nm}, \Re_{m}(\textbf{a}_{-m}))_{m \neq n}, \forall m \in \mathcal{N}\},
    \end{eqnarray}
where $\widetilde{\textbf{X}}_{10}$ is uncertain parameter with uncertainty region as (\ref{unceratintyregioninformation}).
Using the concept of worst-case optimization, the leader's bi-level optimization problem is changed to
\begin{eqnarray}\label{conservativeleader}
  \max_{\textbf{a}_0 \in \mathcal{A}_0} \min_{ \widetilde{\textbf{X}}_{10} \in \mathcal{R}_{\textbf{X}_{10}}}  v_0(\textbf{a}_0,\textbf{f}_0) \\
   \text{subject to:} \quad  \max_{\textbf{a}_{1} \in \mathcal{A}_{1}} \min_{\textbf{f}_1 \in \mathcal{R}_1} u_1(\textbf{a}_{1},\textbf{f}_1). \nonumber
\end{eqnarray}
According to (\ref{conservativeleader}), the leader cannot evaluate precisely its impact on the follower. Note that, 1) the $\textbf{f}_1$ is a linear function of $\textbf{X}_{10}$, and 2) the leader considers the worst-case condition of information in the uncertainty region to obtain the solution of follower, therefore:
\begin{description}
  \item[A5:] for case 2, the negative impact of the leader on the follower is a decreasing function with respect to $\delta_{10}$ i.e., $\nabla_{\delta_{10}}\textbf{f}_1<0$.
\end{description}

\textbf{Remark 2:} The RSE in case 2 always exists, because: 1) $\mathcal{R}_{\textbf{X}_{10}}$, $\mathcal{R}_1$, $\mathcal{A}_0$, and $\mathcal{A}_1$ are compact and closed sets, 2) for any realization of $ \widetilde{\textbf{X}}_{10} \in \mathcal{R}_{\textbf{X}_{10}}$, $\mathcal{R}_1$ is closed and convex. Consequently, for any value of the leader's uncertain information and strategy, the follower has a feasible strategy.

While the existence condition of RSE for case 2 can be shown easily, solving problem (\ref{conservativeleader}) is significantly more complex compared to (\ref{case21}). In the following, we will discuss about the relationship between the RSE for case 2 with the RSE for case 1 and NSE.

\textbf{Proposition 3}: The leader's utility at RSE in case 2 is always less than the leader's utility for case 1 namely,
 $   \omega^{*RSE2}_{0} \leq \omega^{*RSE1}_{0}$,
where $\omega^{*RSE2}_{n}$ is the achieved utility of $n^{\text{th}}$ user in the RSE of case 2.
\begin{proof}
See Appendix D.
\end{proof}
Proposition 3 shows that the leader's incomplete information set always decreases the leader's utility compared to case 1. Next we compare the leader's and follower's utility in the RSE for case 2 to that of the NSE.

\textbf{Proposition 4}: For case 2 of RSG: 1) The leader's strategy is a decreasing function with respect to $\delta_{10}$ and the follower's strategy is an increasing function with respect to $\delta_{10}$, 2) for all realizations of the leader's incomplete information set, we have
\begin{eqnarray}
  \omega_0^{*RSE2} \leq  \omega_0^{*NSE} , \qquad   \omega_1^{*NSE} \leq \omega_1^{*RSE2} \nonumber
\end{eqnarray}
3) the social utility of game increases compared to the NSG i.e., $\omega^{*RSE2}\geq \omega^{*NSE}$, if
\begin{eqnarray}\label{proposition2a}
   \text{C3}:  \quad |\textbf{J}^0_{\textbf{a}_0}| < | \textbf{C}_{10}| \quad , \text{C4}:  \quad  | \textbf{J}^1_{\textbf{a}_1}| > | \textbf{C}_{01}|. \nonumber
\end{eqnarray}
\begin{proof}
Appendix E.
\end{proof}
According to Proposition 4, uncertainty in the leader's information set always decreases the leader's utility compared to NSE. In contrast, the follower reaches a higher utility compared to the NSE. In this case, if the follower's direct rate is greater than its negative impact on the leader (i.e., $\text{C3}$), and the leader's direct rate is less than its negative impact on the follower (i.e., $\text{C4}$), the social utility at RSE for case 2 increases compared to that at NSE.

An interesting interpretation arises when comparing $\text{C1}$ with $\text{C3}$ and $\text{C2}$ with $\text{C4}$. These comparisons indicate that $\text{C1}$-$\text{C2}$ are the dual of $\text{C3}$-$\text{C4}$. In case 1, a higher social utility can be achieved if the increase of the leader's utility is higher than the decares of the follower's utility. In contrast, in case 2, the higher utility can be achieved if the increase of the follower's utility is higher than the decrease of the leader's utility. The variations of utilities of the leader and the follower at RSE compared to NSE for case 1 and 2 are summarized in Table \ref{tablesummery}.

We should note that the implementation of above scenarios in practice does not need the synchronization between leader and follower. For example if the follower plays first, the leader chooses its action based on its information set without considering the follower's action. Then the follower observes the leader's impact and plays based on its observation. Therefore, always the actions of the leader and the follower converge to the NSE and the RSE regardless of synchronization between them.
\section{Illustrative Example}
\begin{figure}
\centering
\includegraphics [height=5cm,width=8cm]{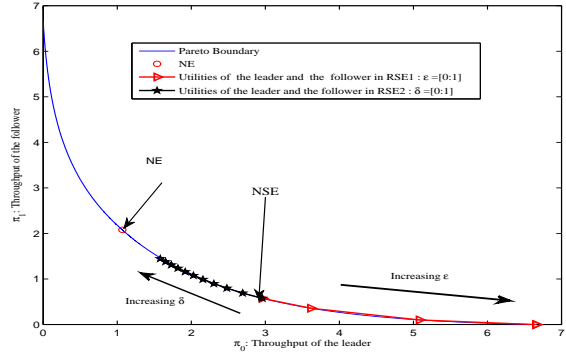}
\caption{Variations of utilities of the follower and the leader for case 1 and case 2 of RSG.}{\label{fig5}}
\end{figure}

In this section, we validate the above results in the power control game and simplify $\text{C1}$-$\text{C4}$ based on channel gains between users. In the power control games, the information set for case 1 equals to $\boldsymbol{\mathcal{I}}^{RSE1}_0=\{(\mathcal{A}_m, v_{m}, \widetilde{\textbf{H}}_{mn}, \textbf{H}_{mm}, \textbf{H}_{nm}, \Re_{m}(\textbf{a}_{-m}))_{m \neq n}, \forall m \in \mathcal{N}\} $. For case 1, the solution of the follower's optimization based on the uncertainty is
\begin{equation}\label{11}
    \widetilde{f}^{*k}_1=f^k_1+\varepsilon_1 \times \frac{\frac{a_{1}^kH_{11}^k}{\sigma^k_1+a_{0}^k H_{01}^k}}{\sqrt{(\sum_{k=1}^K\frac{a_{1}^kH_{11}^k}{\sigma^k_1+a_{0}^k H_{01}^k})^2}} , \quad \forall k \in \mathcal{K},
\end{equation}
First, to illustrate all the results obtained in Proposition 2 and 4, we simulate a single carrier power control using the simulation parameters of the first row in Table \ref{simulation}. Fig. \ref{fig5} shows the variation of utilities of the follower and the leader on the Pareto boundary of the power control game. By increasing $\varepsilon_1$, the leader's utility increases as we expected from Proposition 2. In contrast by increasing the value of $\delta_{10}$, the follower's utility increases and the value of the leader's utility decreases based on Proposition 4.

To provide practical insight into the $\text{C}1$ and $\text{C}2$, we want to express these conditions for power control games only in terms of channel gains. The exact expressions of $\text{C}1$ and $\text{C}2$ for power control game are
\begin{eqnarray}\label{Conditionforpowercontrol}
\text{C1 for power control games}:  \quad \frac{H_{10}^k  H_{11}^{k}a_1^k}{f_1^k \times (f_1^k+H_{11}^{k}a_1^k)}<\frac{H_{00}^k}{f_{0}^k+H_{00}^{k}a_{0}^{k}},  \quad \quad \quad  \forall k \in \mathcal{K},\\
\text{C2 for power control games}:  \quad \frac{ H_{01}^k H_{00}^{k}a_0^k}{f_0^k \times (f_0^k+H_{00}^{k}a_0^k)}>\frac{H_{11}^k}{f_{1}^k+H_{11}^{k}a_{1}^{k}}, \quad \quad \quad \forall k \in \mathcal{K}.
 \end{eqnarray}
To simplify the above conditions, we consider three scenarios based on signal-to-interference-plus-noise ratio (SINR) of the leader and the follower as: R1) High SINR scenario i.e., $ H_{00}^{k}a_{0}^k \gg H_{01}^{k}a_{1}^k+\sigma_{0}^{k}$ and $H_{11}^{k}a_{1}^k \gg H_{10}^{k}a_{0}^k+\sigma_{1}^{k}$ in which $\text{C1}$ and $\text{C2}$ are simplified to
\begin{equation}\label{HighSIR}
    H^k_{10}<H_{01}^{k},
\end{equation}
R2) Low SINR scenario i.e., $H_{00}^{k}a_{0}^k \ll H_{01}^{k}a_{1}^k+\sigma_{0}^k$ and $H_{11}^{k}a_{1}^k \ll H_{10}^{k}a_{0}^k+\sigma_{1}^k$, where the social utility increases if
\begin{equation}\label{lowsir}
    H_{00}^k>H_{01}^k \quad  \text{and} \quad H_{11}^k<H_{10}^k,
\end{equation}
and, R3) Moderate SINR scenario, i.e., where the values of induced interference of the leader and the follower to each other are close, i.e., $f^k_1 \approx f_0^k$. Therefore, $\text{C1}$ and $\text{C2}$ change to
\begin{eqnarray}\label{C1aforpowercontrolR3}
    \frac{H^k_{00}}{H_{01}^k}> \frac{H_{11}^k a_{1}^k}{H_{10}^{k} a_{0}^{k}+\sigma_{1}^{k}} \quad ,  \quad  \frac{H_{11}^{k}}{H_{10}^k} < \frac{H_{00}^k a_{0}^k}{H_{01}^k a_{1}^{k}+\sigma_{0}^{k}} \quad \forall k \in \mathcal{K},  \nonumber
\end{eqnarray}
The above conditions are simplified by assuming that channel noise is much less than the interference from other users, e.g., $H_{10}^{k} a_{0}^{k}\gg \sigma_{1}^{k}$, and the power transmissions of the leader and the follower are close, as
\begin{eqnarray}\label{C1aforpowercontrolsimlified}
    H^k_{00}H_{10}^k> H_{11}^k H_{10}^k  \quad \quad\quad \forall k \in K.
\end{eqnarray}
Using (\ref{HighSIR})-(\ref{C1aforpowercontrolsimlified}), we can predict how the social utility increases or decreases under given channel conditions for case 1. To provide insight about implementing the Stackelberg game in the interference channels, we investigate the effect of these conditions on the amount of decrease and increase of the leader's utility and follower's utility. Consider $d_{n}^{RSE1}=\frac{\omega_{n}^{RSE1}-\omega_{n}^{NSE}}{\omega_{n}^{NSE}}$ to show the percentage of change of utility of player $n$ in RSE for case 1 compared to NSE, and $d^{RSE1}=\frac{\omega^{RSE1}-\omega^{NSE}}{\omega^{NSE}}$ as percentage of change of social utility of RSE for case 1 compared to NSE. A larger value of $d_{n}^{RSE1}$ indicates a larger increase of utility of player $n$ in case 1. The same notation for case 2 will be applied as $d_{n}^{RSE2}=\frac{\omega_{n}^{RSE2}-\omega_{n}^{NSE}}{\omega_{n}^{NSE}}$ and $d^{RSE2}=\frac{\omega^{RSE2}-\omega^{NSE}}{\omega^{NSE}}$ for user $n$ and social utility, respectively.

\begin{figure}
\centering
\includegraphics [height=8.cm,width=9.25cm]{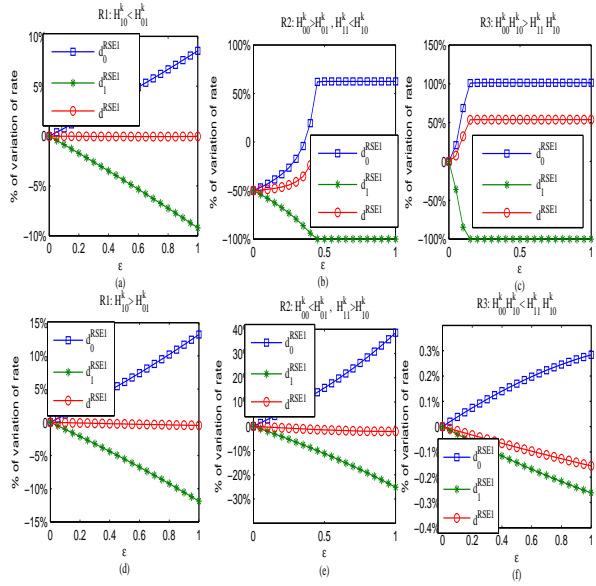}
\caption{Validating Proposition 2 with the numerical example of power control. (a) in R1 and (\ref{HighSIR}) holds (b) in R2 and (\ref{lowsir}) holds, (c) in R3 and (\ref{C1aforpowercontrolsimlified}) holds, (d) in R1 and (\ref{HighSIR}) does not hold (e) in R2 and (\ref{lowsir}) does not hold, (g) in R3 and (\ref{C1aforpowercontrolsimlified}) does not hold.} {\label{Figcase1}}
\end{figure}

In Figs. \ref{Figcase1} (a), (b) and (c), we depict $d_{0}^{RSE1}$, $d_{1}^{RSE1}$ and $d^{RSE1}$ for R1, R2, and R3, respectively. Increasing the leader's utility and decreasing the follower's utility are evident with respect to $\varepsilon_1$ as expected from Proposition 2. When (\ref{HighSIR})- (\ref{C1aforpowercontrolsimlified}) are satisfied, the social utility increases as shown in Fig. \ref{Figcase1} (a), (b), and (c). In contrast, when (\ref{HighSIR})- (\ref{C1aforpowercontrolsimlified}) are not satisfied, the social utility decreases compared to NSE as depicted in Figs. \ref{Figcase1} (e), (f) and (g). However, in R1, the increase of social utility is not considerable if (\ref{HighSIR}) holds. In R2 and R3, when (\ref{lowsir}) and (\ref{C1aforpowercontrolsimlified}) are satisfied, the leader's utility and social utility increase considerably. For example, when (\ref{lowsir}) does not hold, $d_0^{RSE1}$ reduces from $220\%$ in Fig. \ref{Figcase1}. (b) to $40\%$ in Fig. \ref{Figcase1}. (e), and $d^{RSE1}$ decreases from $100\%$ to around $-1\%$. Therefore, both of the accurate information set of leader and holding the (\ref{lowsir}) are valuable from leader and system.  The same is true for R3.

\begin{figure}
\centering
\includegraphics [height=8.cm,width=9.25cm]{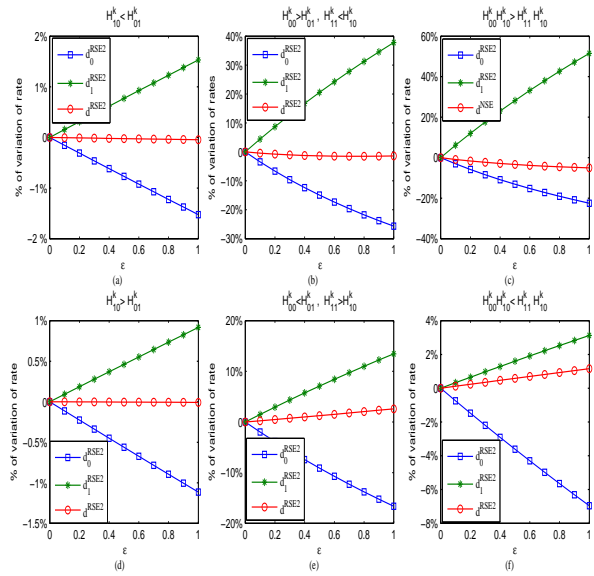}
\caption{ Validating Proposition 4 with power control games. (a) in R1 and (\ref{highsir2}) does not hold, (b) in R2 and (\ref{lowsir2}) does not hold, (c)  in R3 and (\ref{moderate2}) does not hold, (d) in R1 and (\ref{highsir2}) holds, (e) in R2 and (\ref{lowsir2}) holds, (g) in R3 and (\ref{moderate2}) holds.} {\label{Figcase2}}
\end{figure}

For case 2, the uncertain parameter is $\textbf{H}_{10}$. For the above regions, $C_3$ and $C_4$ change to
\begin{eqnarray} \label{highsir2}
  \text{R1:}  &\Longrightarrow& H^k_{10}>H_{01}^{k} \\
  \text{R2:} &\Longrightarrow&  H_{00}^k<H_{01}^k , \quad H_{11}^k>H_{10}^k \label{lowsir2}\\
 \text{R3:} &\Longrightarrow& H^k_{00}H_{10}^k<H_{11}^k H_{10}^k  \forall k.  \label{moderate2}
\end{eqnarray}
The effects of enlarging the value of $\delta_{01}$ in $d_{0}^{RSE2}$, $d_{1}^{RSE2}$ and $d^{RSE2}$ are investigated in Figs. \ref{Figcase2} (a)-(f). As we expected from Proposition 4, the leader's utilities in all the cases decrease compared to that at the NSE, while the follower's utility increases by increasing the value of $\delta_{01}$. In Figs. \ref{Figcase2} (a), (b), and (c), conditions (\ref{highsir2})-(\ref{moderate2}) do not hold. Consequently, the social utility decreases. In contrast, in Figs. \ref{Figcase2}. (d), (e), and (f), the social utility increases by increasing $\delta_{10}$ because (\ref{highsir2})-(\ref{moderate2}) hold. Fig. \ref{Figcase2} (a) and \ref{Figcase2} (d) show that $d_{1}^{RSE2}$ decreases from $1.5\%$ to $1\%$ when (\ref{highsir2}) holds compare to the case it dose not hold. The same decreasing is observable by comparing $d_{1}^{RSE2}$ in Figs \ref{Figcase2} (b) and \ref{Figcase2} (e) and Fig. \ref{Figcase2} (c) and (f). These comparisons indicate that in all the regions, the follower obtains a lower utility when (\ref{highsir2})-(\ref{moderate2}) hold and the social utility increases insignificantly.
\subsection{Practical remarks}
Comparing Fig. \ref{Figcase1} and Fig. \ref{Figcase2} shows that the leader's utility and social utility increase significantly for case 1. In contrast for case 2, the social utility has lower rate of increase. Also the follower's utility increases insignificantly. For example, in R2, in case 1 the leader's utility and social utility increase up to $200\%$, while for case 2, the follower's utility and social utility increases up to $10\%$ in R2. Therefore, the leader's complete information set is more effective in increasing the social utility. This analysis provides incentives for the system designer to build more efficient protocols to collect the accurate information for the leader. Also these results help in designing systems with two or more leaders. Consider that leaders want to increase their social utility and they encounter uncertainty in their parameters. If (\ref{HighSIR})- (\ref{C1aforpowercontrolsimlified}) hold for one of them and it has a complete information, it plays as a leader and the others play as followers. Therefore the social utility of leaders increases considerably.

\subsection{Power control with maximum transmit power constraints}
\begin{figure}
\centering
\includegraphics [height=6.5cm,width=8cm]{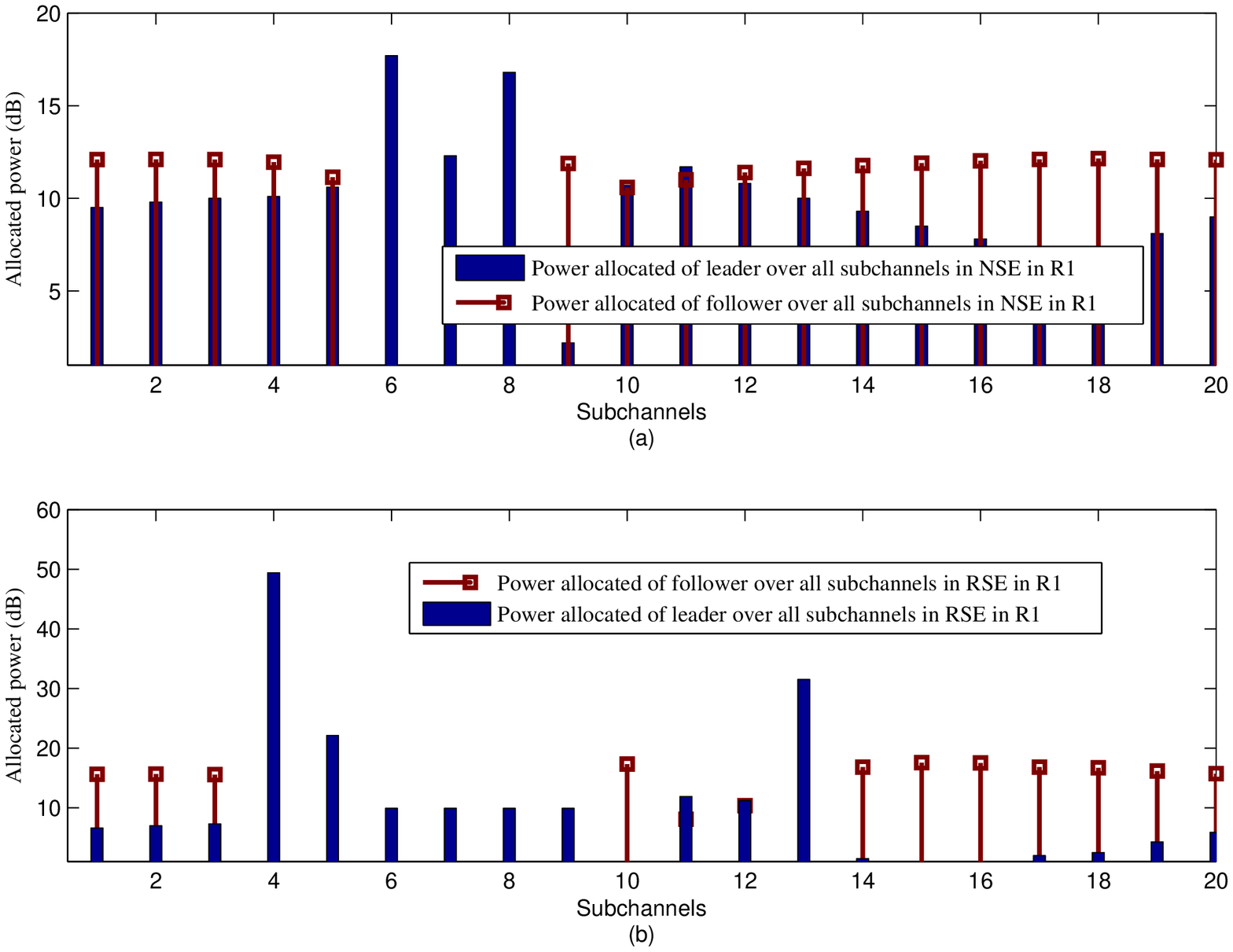}
\caption{ Power allocation of the follower and the leader in R1 subject to (\ref{powerlimitation}) (a) at NSE and (b) at RSE for case 1.} {\label{limittedpower1}}
\end{figure}

\begin{figure}
\centering
\includegraphics [height=7cm,width=8cm]{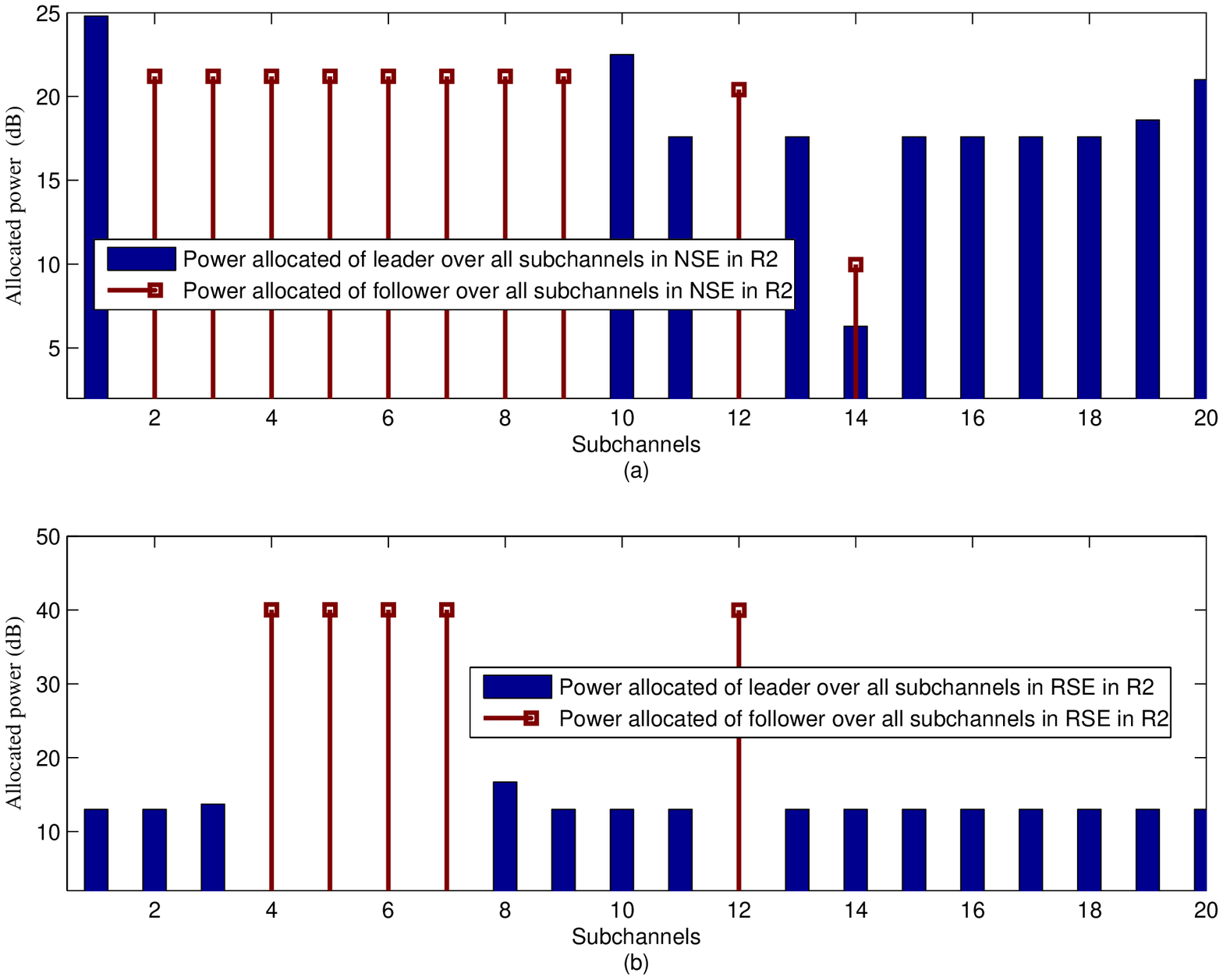}
\caption{ Power allocation of the follower and the leader in R2 subject to (\ref{powerlimitation}) (a) at NSE and (b) at RSE for case 1.} {\label{limittedpower2}}
\end{figure}

\begin{figure}
\centering
\includegraphics [height=7cm,width=8cm]{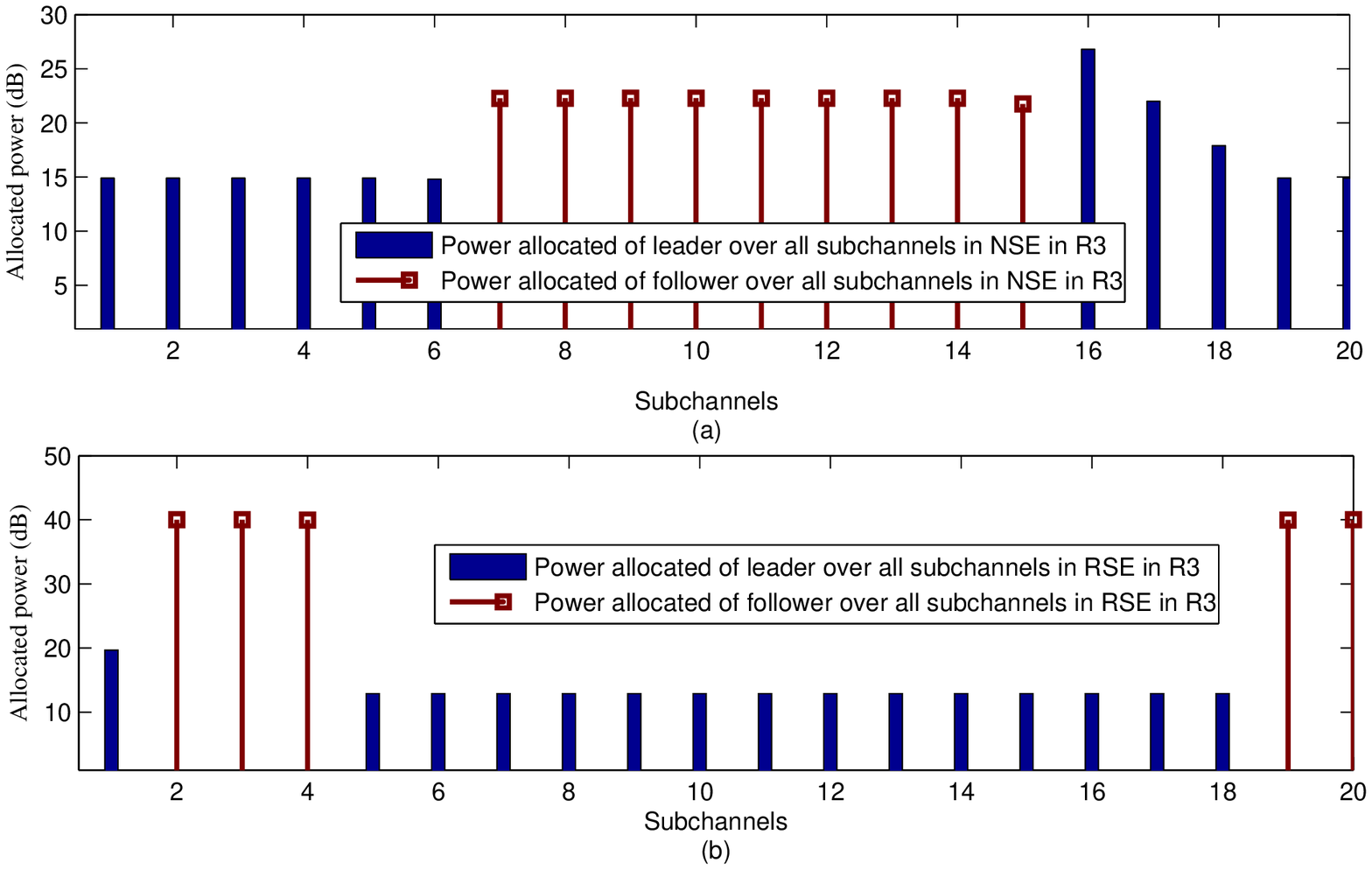}
\caption{ Power allocation of the follower and the leader in R2 subject to (\ref{powerlimitation}) (a) at NSE and (b) at RSE for case 1.} {\label{limittedpower3}}
\end{figure}

Consider that the sum of powers of each player over all subchannels is limited by $P_n^{\text{max}}$, i.e.,
\begin{equation}\label{powerlimitation}
    \sum_{k=1}^{K} a_{n}^{k}\leq P_n^{\text{max}}.
\end{equation}
In this case, the players' strategies are nonlinear functions with respect to their observations \cite{newprespetive,closedformsolution}. Therefore, we cannot use Propositions 1-4 directly. To investigate the performance of RSE with (\ref{powerlimitation}), we show the allocated power over different subchannels in Figs. \ref{limittedpower1}-\ref{limittedpower3}. The corresponding leader's utility and follower's utility are summarized in Table \ref{Tableutilitypower}. The simulation parameters are the same as in Figs. \ref{Figcase1} - \ref{Figcase2} in Table \ref{simulation} except $P_n^{\text{max}}=a_{nk}^{\text{max}}=200$ dB.

Consider $\mathcal{K}^{NSE}_n\subseteq \mathcal{K}$ and $\mathcal{K}^{RSE1}_n\subseteq \mathcal{K}$ as the sets of subchannels utilized by user $n$ at the NSE and RSE for case 1, respectively. Also we define $\mathcal{L}^{NSE}_{nm}=\mathcal{K}^{NSE}_n  \cap \mathcal{K}^{NSE}_m$ and  $\mathcal{L}^{RSE1}_{nm}=\mathcal{K}^{RSE1}_n  \cap \mathcal{K}^{RSE1}_m$ as the set of common subchannels between user $m$ and user $n$ at the NSE and the RSE for case 1, respectively. According of the simulation results in Figs. \ref{limittedpower1}-\ref{limittedpower3}, we have
$ |\mathcal{L}^{RSE1}_{01}| < |\mathcal{L}^{NSE}_{01}|$, where $|\mathcal{L}^{RSE1}_{01}|$ and  $|\mathcal{L}^{NSE}_{01}|$ are the size of $\mathcal{L}^{RSE1}_{01}$ and $\mathcal{L}^{NSE}_{01}$, respectively.
For example, $|\mathcal{L}^{RSE1}_{01}|=17$ and  $|\mathcal{L}^{NSE}_{01}|=13$ in R1 and $ |\mathcal{L}^{RSE1}_{01}|=0$ and  $|\mathcal{L}^{NSE}_{01}|=1$ in R2. As we expected and seen in Table \ref{Tableutilitypower}, the leader's utility increases for case 1 in all conditions. Interestingly, in R2, the follower's utility also increases compared to NSE. Since the number of common subchannels decrease, there is  less interference from the leader to the follower and viceversa at RSE for case 1 compared to NSE. Therefore, as shown in Table \ref{simulation}, there is a probability that the leader's utility and follower's utility increase simultaneously.

To quantify this probability, we show the simulated cumulative distributed function (cdf) of $d_1^{RSE1}$ for 3 situations of high interference scenarios between the leader and the follower as: S1) the leader's interference to the follower is high, e.g., $\frac{H_{10}^k}{H_{11}^k}>0.8$, but the follower's interference to the leader is low, e.g., $\frac{H_{01}^k}{H_{00}^k}<0.1$; S2) the leader's and the follower's interference are high to each other, e.g., $\frac{H_{10}^k}{H_{11}^k}>0.9$ and $\frac{H_{01}^k}{H_{00}^k}>0.9$; S3) the follower's interference to the leader is high; e.g., $\frac{H_{10}^k}{H_{11}^k}<0.1$; but for the leader to the follower is low, e.g., $\frac{H_{01}^k}{H_{00}^k}>0.9$.

\begin{figure}
\centering
\includegraphics [height=6.5cm,width=8cm]{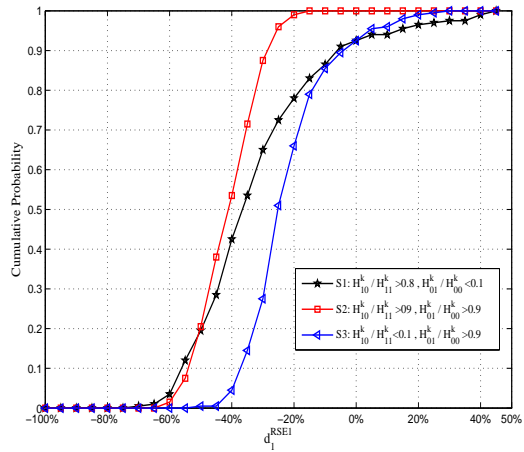}
\caption{ CDF of $d_{1}^{RSE1}$ when (\ref{powerlimitation}) holds.} {\label{pf}}
\end{figure}

From the simulated curves in Fig. \ref{pf}, when both the interference from the leader to the follower and the follower to the leader are large, i.e., in S2 there is no probability for the follower to reach a higher utility compared to NSE. In contrast, in S1 and S3, there is a probability that the  follower's utility increases e.g., around $0.1$. Therefore, the leader and the follower reach higher utilities when there is uncertainty in the follower's observations with power constraint in (\ref{powerlimitation}) which is an opportunistic phenomena of RSG in case 1 in terms of increasing utility of both leader and follower.
\section{Extension to Multi-user Game}
The pervious results can be extended for multiple-leaders and multiple-followers in the Stackelberg game. Obviously, analysis of RSE in this scenario is more challenging \cite{Mihaealalinearly,Generalizednash,GlobalOptimization}. In order to do that, we focuss on the case that the NE of game of followers is unique.
\subsection{One leader and multiple-followers ($N_l=1$ and $N_f>1$) } For this scenario there is only one leader in the Stackelberg game denoted with the index $0$. Consider the $N_f \times N_f$ matrix $\boldsymbol{\Upsilon}$ by following elements
\begin{eqnarray}
   [ \Upsilon]_{nm}= \left\{\begin{array}{c}
  \alpha_{n}^{\text{min}} \,\qquad\text{if} \qquad\, m=n, \, \, m, n \in \mathcal{N}_f\nonumber \\
-\beta_{nm}^{\text{max}}  \,\qquad  \text{if} \qquad\, m\neq n,  \, m, n \in \mathcal{N}_f \nonumber \\
\end{array} \right.
\end{eqnarray}
where
\begin{eqnarray}
  \alpha_{n}(\textbf{a})\triangleq \text{smallest eigenvalue of}
    \quad  -\nabla^{2}_{\textbf{a}_{n}} v_{n}(\textbf{a}_n, \textbf{f}_n) \quad \alpha_{n}^{\text{min}}\triangleq \inf_{\textbf{a} \in \mathcal{A}}  \alpha_{n}(\textbf{a}) \,\,\quad \forall n \in \mathcal{N}_f \\
  \beta_{nm}(\textbf{a})\triangleq\|- \nabla_{\textbf{a}_{n}\textbf{a}_{m}}v_{n} (\textbf{a}_n, \textbf{f}_n) \|
    \quad \forall n\neq m \quad \beta_{nm}^{\text{max}}\triangleq \sup_{\textbf{a} \in \mathcal{A}}
    \beta_{n}(\textbf{a}) \,\,\quad \forall n \in \mathcal{N}_f
 \end{eqnarray}
If $\boldsymbol{\Upsilon}$ is a $P$-matrix, the NE between followers is unique (Theorem
12.5, \cite{palomar2010}).
\subsubsection{Case 1 of RSE}
In this case, the followers' observations are uncertain parameters which are modeled as (\ref{III-1}). In contrast, the leader has the complete information set. For each follower, the optimization problem is similar to (\ref{utilityrobustfollower}) and the reformulations to (\ref{vertana}) can be applied as
 \begin{equation}\label{followerstartegyspace}
  \widetilde{\textbf{f}}^{*}_{n}=\textbf{f}_n- \varepsilon_n \boldsymbol{\vartheta}_{n}, \, \quad \quad n \in \mathcal{N}_f
\end{equation}
where $\widetilde{\textbf{f}}_n^*=[\widetilde{f}^{1*}_{n},\cdots, \widetilde{f}^{K*}_{n}]$, $\boldsymbol{\vartheta}_{n}=[\vartheta^1_{n}, \cdots, \vartheta^K_{n}]$, and
\begin{equation}\label{vertanamultiplePU}
    \vartheta^k_{n} =\frac{\frac{\partial u^k_n(\textbf{a}_n, \widetilde{\textbf{f}}_n)}{\partial f_n^k}}{\sqrt{\sum_{k=1}^{K} (\frac{\partial u^k_n(\textbf{a}_n, \widetilde{\textbf{f}}_n)}{\partial f_n^k})^2}}.
\end{equation}
\textbf{Proposition 5}: For case 1 of RSG, if $\boldsymbol{\Upsilon}$ is a $P$-matrix; 1) the followers' strategies are decreasing functions with respect to $\boldsymbol{\varepsilon}=[\varepsilon_1,\cdots, \varepsilon_N]$ and the social utility of followers' game is less than that of the NSE; 2) the leader's utility at RSE is higher than that at the NSE; 3) the social utility increases compared to NSE if
\begin{eqnarray}
\text{C5}: \, \textbf{J}^0_{\textbf{a}_0}> \sum_{n \in \mathcal{N}_f} \textbf{C}_{n 0} \, , \quad \quad \text{C6}: \,\,  \textbf{J}^n_{\textbf{a}_n}< \textbf{C}_{0n} + \sum_{m\neq n, m \in \mathcal{N}_f} \textbf{C}_{mn}, \quad \forall n \in \mathcal{N}_f. \nonumber
  \end{eqnarray}
\begin{proof}
See Appendix F.
\end{proof}
The same as in  Proposition 2, the followers' noisy observations increase the leader's utility. In contrast, they decrease the followers' utilities. The social utility increases in RSE compared to that at NSE, if the leader's direct rate is larger than its negative impact on the followers i.e., $\text{C5}$, and, for each follower, its negative impacts on the other followers and the leader are greater than on its direct rate i.e., $\text{C6}$. Therefore under these conditions, decreasing  the follower's actions has more effect to increase of the leader's utility and the other followers' utilities compared to decrease of its utility. Also, increase of the leader's action has more effect to increase of its utilities compared to decrease of the followers' utilities.
\subsubsection{Case 2} In this case, we assume that $\textbf{X}_{n0}$ is uncertain parameter and the leader considerers the uncertainty region with boundary $\delta_{n0}$ for each follower as
\begin{equation}\label{unceratintyregioninformation2}
    \mathcal{R}_{\textbf{X}_{n0}}=\{ \widetilde{\textbf{X}}_{n0} \, | \,\, \| \widehat{\textbf{X}}_{n0} \|_2= \|\widetilde{\textbf{X}}_{n0}- \textbf{X}_{n0}\|_2 \leq \delta_{n0} \}, \quad \forall n \in \mathcal{N}_f.
\end{equation}
\textbf{Proposition 6}: For case 2, if $\boldsymbol{\Upsilon}$ is a $P$-matrix, 1) the leader's utility is always less than that of the NSE;
 2) the followers' actions are increasing functions with respect to $\boldsymbol{\delta}_0=[\delta_{10},\cdots, \delta_{N_f,0}]$, and the social utility of the followers' game is higher than that of the NSE; 3) the social utility increases if
\begin{eqnarray}
\text{C7}: \, \textbf{J}^0_{\textbf{a}_0}< \sum_{n \in \mathcal{N}_f} \textbf{C}_{n 0}, \,\, \quad \quad \text{C8}: \, \textbf{J}^n_{\textbf{a}_n}> \textbf{C}_{0n} + \sum_{m\neq n, m \in \mathcal{N}_f} \textbf{C}_{mn}, \forall n \in \mathcal{N}_f. \nonumber
  \end{eqnarray}
\begin{proof}
See Appendix G.
\end{proof}
Again, the leader's incomplete information set decreases the leader's utility and increases the social utility of followers. Also from $\text{C}7$-$\text{C}8$, if the leader's direct rate is less than its negative impact on the followers and negative impacts of each follower on the other followers and the leader are less than of its direct rate, the social utility of RSE for case 2 increases compared to NSE. Besides, $\text{C}5$- $\text{C}6$, and $\text{C}7$-$\text{C}8$ are dual.

For multiple-followers and one leader scenario, the hierarchy between the leader and followers still remains. To implement this scenario in a distributed manner, the leader announces its action first. Then, all the followers play the strategic non-cooperative robust game. The robust game between followers can be implemented using a distributed algorithm as in \cite{saeedeh5}.
\subsection{Multiple-leaders and Multiple-followers: $N_l>1$ and $N_f>1$}
In this scenario, the definition of RSG and RSE is not unique. There are different forms of NSG for multiple-leaders and multiple-followers scenario based on interaction among the leaders such as cooperation or competition \cite{Generalizednash,NEexistence}. For example, one of the possible way to define the Stackelberg game with multiple-followers is to consider cooperation between leaders when all leaders know each other and try to maximize their social utility as
\begin{eqnarray}\label{multileadersccenario}
   &\max_{ \textbf{a}_n \in \mathcal{A}_n} \sum_{n \in \mathcal{N}_l}& v_n(\textbf{a}_n, \textbf{f}_n) \\
   &\text{subject to}& \max_{ \textbf{a}_m \in \mathcal{A}_m}  v_m(\textbf{a}_m, \textbf{f}_m) \quad \quad \forall m \in \mathcal{N}_f, \nonumber
\end{eqnarray}
\textbf{Proposition 7:} If $\boldsymbol{\Upsilon}$ is a P-matrix for the game of followers, leaders' social utility increases for case 1 of RSE.
\begin{proof}
See Appendix H.
\end{proof}
In this scenario, the major concern is the analysis of case 2. Since (\ref{multileadersccenario}) is a non-convex and non-smooth optimization problem, deriving the conditions for increasing or decreasing the aggregate of leaders' utilities (\ref{multileadersccenario}) is impossible. Nevertheless, a heuristic protocol to increase the social utility of followers and some of the leaders is proposed based on the results of this paper in section IV. This protocol is summarized in Table \ref{table4}. Briefly, in this protocol, the leader with the highest negative impact on the followers and the other leaders and the less internal rate is chosen to play as the leader and all the others leaders and followers are considered as the followers. Therefore, the social utility of game among the followers and the remaining leaders increases with respect to the leader's incomplete information set based on Proposition 6.

In Table \ref{tablehuresticalgorrithm}, we evaluate the performance of heuristic algorithm where there are two leaders and only one follower in power control game. In SNE, both of the leaders maximize their sum of utility based on (\ref{multileadersccenario}). When the information of leaders are subject to uncertainty, leader 1 acts as the leader and leader 2 as the follower. In this case, the uncertainty in the leaders causes the decreasing utilities of both of the leaders, while the utility of follower increases. However, the social utility of the follower and the leader 2 increases for the case 2 of RSE with heuristic algorithm. Clearly, the above heuristic algorithm does not guarantee to increase the leader's social utility. However, it guarantees that social utility over all players excepts player $n_l$ increases for case 2 of RSE.
\section{Conclusion}
In this paper, we introduced the robust Stackelberg equilibrium for communication systems under different types of uncertainty. We derive that the followers' uncertain information increases the leaders' utilities while decreasing the followers' utilities and  the leaders' uncertain information increases the followers' utilities while decreasing the leaders' utilities. We show that the social utility in case 1 increases if the increasing rates of the leaders' utilities are larger than the decreasing rates of the followers' utilities. In contrast, in case 2, if the decreasing rates of the leaders' utilities are less than the increasing rates of the followers' utilities, the social utility increases. For power control games, we simplify these general conditions based on the channel gains between transmitter and receivers in different SINR scenarios and provide insights on how to design enhanced protocol for gathering the leaders' information and implement high-efficient systems in the multiple-leaders communication scenario. These results can be extended to other communication systems to gain insights on how to implement the RSE and NSE.
\bibliographystyle{IEEEtran}
\bibliography{IEEEabrv,mybib7}

\begin{thebibliography}{10}
\providecommand{\url}[1]{#1}
\csname url@samestyle\endcsname
\providecommand{\newblock}{\relax}
\providecommand{\bibinfo}[2]{#2}
\providecommand{\BIBentrySTDinterwordspacing}{\spaceskip=0pt\relax}
\providecommand{\BIBentryALTinterwordstretchfactor}{4}
\providecommand{\BIBentryALTinterwordspacing}{\spaceskip=\fontdimen2\font plus
\BIBentryALTinterwordstretchfactor\fontdimen3\font minus
  \fontdimen4\font\relax}
\providecommand{\BIBforeignlanguage}[2]{{%
\expandafter\ifx\csname l@#1\endcsname\relax
\typeout{** WARNING: IEEEtran.bst: No hyphenation pattern has been}%
\typeout{** loaded for the language `#1'. Using the pattern for}%
\typeout{** the default language instead.}%
\else
\language=\csname l@#1\endcsname
\fi
#2}}
\providecommand{\BIBdecl}{\relax}
\BIBdecl

\bibitem{basar}
T.~Basar and G.~J. Olsder, \emph{Dynamic Non-Cooperative Game Theory},
  2nd~ed.\hskip 1em plus 0.5em minus 0.4em\relax Academic press, New yourk,
  Feb. 1995.

\bibitem{NEexistence}
D.~Fudenberg and J.~Tirole, \emph{Game Theory}.\hskip 1em plus 0.5em minus
  0.4em\relax Cambridge, MA: MIT Press, 1991.

\bibitem{newprespetive}
Y.~Su and M.~van~der Schaar, ``A new perspective on multi-user power control
  games in interference channels,'' \emph{IEEE Trans. Wireless Commun.},
  vol.~8, no.~6, pp. 2910--2919, Jun. 2009.

\bibitem{knowledgeandlearning}
------, ``Decentralized knowledge and learning in strategic multi-user
  communication,'' {2011. [Online]. Available:http://arxiv.org/abs/0804.2831.}

\bibitem{stackelburggame}
L.~Lai and H.~E. Gamal, ``The water-filling game in fading multiple-access
  channels,'' \emph{IEEE Transactions On Information Theory}, vol.~54, no.~5,
  pp. 2110--2122, May 2008.

\bibitem{Gershman}
A.~B. Gershman and N.~D. Sidiropoulos, \emph{Space-time processing for {MIMO}
  communications}.\hskip 1em plus 0.5em minus 0.4em\relax John Wiley and Sons,
  2005.

\bibitem{selecectedrobust}
A.~Ben-Tal and A.~Nemirovski, ``Selected topics in robust convex
  optimization,'' \emph{Mathematical Programming}, vol.~1, no.~1, pp.
  125–--158, July 2007.

\bibitem{Generalnorm}
D.~Bertsimas, D.~Pachamanovab, and M.~Sim, ``Robust linear optimization under
  general norms,'' \emph{Operations Research Letters}, vol.~4, no.~32, pp.
  510--516, 2004.

\bibitem{Bental2000}
A.~Ben-Tal and A.~Nemirovski, ``Robust solutions to uncertain programs,''
  \emph{Operations Research Letter}, vol.~25, p. 1–13, Feb. 1999.

\bibitem{Stochasticstack}
T.-S. Chang and Y.-C. Ho, ``Stochastic stackelberg games: Nonnested multistage
  multiagent incentive problems,'' \emph{IEEE Transactions on Automatic
  Control}, vol.~28, no.~4, pp. 477 -- 488, Apr. 1983.

\bibitem{ffoundded}
M.~Anam, S.~A. Basher, and S.~H. Chiang, ``Mixed oligopoly under demand
  uncertainty,'' \emph{Journal of Theoretical Economics Topics}, vol.~7, no.~1,
  2007.

\bibitem{Incentivesdesign}
C.~Xu and K.~Kijima, ``Incentives design under parametric uncertainty,''
  \emph{IEEE Transactions on Systems Man and Cybernetics Part A: Systems and
  Humans}, vol.~28, no.~3, pp. 339 -- 346, May 1998.

\bibitem{Localrobustness}
W.~A. Brock and S.~N. Durlauf, ``Local robustness analysis: Theory and
  application,'' \emph{Elsevier Journal of Economic Dynamics and Control},
  vol.~29, no.~11, pp. 2067--2092, Nov. 2005.

\bibitem{Bassarincetive}
D.~H. Cansever and T.~Basar, ``A minimum sensitivity approach to incentive
  design problems,'' \emph{Large Scale Systems}, vol.~5, pp. 233--244, 1983.

\bibitem{CP}
M.~Chiang, P.~Hande, T.~Lan, and C.~W. Tan, ``Power control in wireless
  cellular networks,'' \emph{Foundations and Trends in Networking}, vol.~2,
  no.~4, pp. 381--533, July. 2008.

\bibitem{ResourceAllocationinWirelessNetworks}
S.~St{a´}nczak, M.~Wiczanowski, and H.~Boche, \emph{Resource Allocation in
  Wireless Networks: Theory and Algorithms}.\hskip 1em plus 0.5em minus
  0.4em\relax Berlin, Germany: Springer, 2006.

\bibitem{mihaelastructure}
Y.~Su and M.~van~der Schaar, ``Structural solutions for additively coupled sum
  constrained games,'' \emph{IEEE Trans. on Information Theory.
  \text{(Submitted)}}, [Online]. Available:arxiv.org/abs/1005.0880.

\bibitem{ref1}
E.~Altman, T.~Boulogne, R.~El-Azouzi, T.~Jimenez, and L.~Wynter, ``A survey on
  networking games in telecommunications,'' \emph{Computer Operation Research},
  vol.~33, no.~2, pp. 286--311, Feb. 2006.

\bibitem{Convergenceofiterativewaterfillingalgorithm}
K.~W. Shum, K.-K. Leung, and C.~W. Sung, ``Convergence of iterative
  waterfilling algorithm for gaussian interference channels,'' \emph{IEEE J.
  Sel. Areas Commun.}, vol.~25, no.~6, pp. 1091--1100, Aug. 2007.

\bibitem{bilevel}
B.~Colson, P.~Marcotte, and G.~Savard, ``Bilevel programming: A survey,''
  \emph{A quarterly Journal of Operation Research}, vol.~3, no.~2, pp. 87--107,
  2005.

\bibitem{saeedeh5}
S.~Parsaeefard, A.~R. sharafat, and M.~van~der Schaar, ``Robust additivly
  coupled game,'' 2011 [Online]. Available:arxiv.org/abs/0212325.

\bibitem{Robustgame}
M.~Aghassi and D.~Bertsimas, ``Robust game theory,'' \emph{Math. Program.}, pp.
  231–--273, 2006.

\bibitem{wosrtcasepalomar}
J.~Wang and D.~P. Palomar, ``Worst-case robust {MIMO} transmission with
  imperfect channel knowledge,'' \emph{IEEE Trans. Signal Proc.}, vol.~57,
  no.~8, pp. 3086--3100, Aug. 2009.

\bibitem{closedformsolution}
E.~Altman, K.~Avrachenkov, and A.~Garnaev, ``Closed form solutions for
  water-filling problems in optimization and game frameworks,'' \emph{Spring
  Journal of Telecommunication Systems}, vol.~47, no. 1-2, pp. 153--164, 2011.

\bibitem{Mihaealalinearly}
Y.~Su and M.~van~der Schaar, ``Linearly coupled communication games,''
  \emph{IEEE Trans. Commun., accpted to publish}, {2011. [Online]. Available:
  http://arxiv.org/abs/0908.1613v1.pdf}.

\bibitem{Generalizednash}
J.-S. Pang and M.~Fukushima, ``Quasi-variational inequalities, generalized nash
  equilibria, and multi-leader-follower games,'' \emph{Computational Management
  Science}, vol.~2, no.~1, pp. 21--56, 2005.

\bibitem{GlobalOptimization}
Z.~H. Gumus and C.~A. Floudas, ``Global optimization of nonlinear bilevel
  programming problems,'' \emph{Journal of Global Optimzation}, vol.~20, no.~1,
  pp. 1–--31, 2001.

\bibitem{palomar2010}
D.~P. Palomar and Y.~C. Eldar, \emph{Convex optimization in signal processing
  and communications}.\hskip 1em plus 0.5em minus 0.4em\relax Cambridge
  university press, 2010.

\bibitem{boydconvexbook}
S.~Boyd and L.~Vandenberghe, \emph{Convex Optimization}.\hskip 1em plus 0.5em
  minus 0.4em\relax Cambridge University Press, 2004.

\end{thebibliography}
\appendices
\section{Proof of Lemma 1}
Since $u_1(\textbf{a}_1, \textbf{f}_1)$ is convex with respect to $\textbf{f}_1$, the  internal optimization problem of the follower in (\ref{case21}) can be solved as
 \begin{equation}\label{lagrangedualfunction}
    L(\textbf{a}_1,\widetilde{\textbf{f}}_{1}, \lambda)= \sum_{k=1}^{K} u^{k}_1(a_1^k,\widetilde{f}^{k}_{1}) + \lambda ( \sum_{k=1}^{K} (\widetilde{f}_{1}^{k}-f_{1}^{k})^2- \varepsilon_1^2),
 \end{equation}
where $\lambda$ is the nonnegative Lagrange multiplier that satisfies (\ref{III-1}), i.e.,
  \begin{equation}\label{lagrangemultiplier}
\lambda \times (\varepsilon_1^2 - \sum_{k=1}^{K} (\widetilde{f}_{1}^{k}-f_{1}^{k})^2)=0.
 \end{equation}
The solution of (\ref{lagrangedualfunction}) according to $\widetilde{f}_{1}^{k}$ can be obtained by optimality condition of optimization problem without constraint \cite{boydconvexbook}, i.e., $\frac{\partial L(\textbf{a}_1,\boldsymbol{\widetilde{f}}_{1}, \lambda)}{\partial \widetilde{f}^{k}_1}=0$, which is equivalent to
\begin{equation}\label{solution}
    \frac{\partial u^k_1(a^k_{1},\widetilde{f}^k_{1})  }{\partial \widetilde{f}^k_{1}}= -2 \lambda \times (\widetilde{f}_1^k-f_1^k)  \qquad \forall k \in \mathcal{K}.
\end{equation}
If the above solution is inserted in (\ref{lagrangemultiplier}), the uncertain parameter is obtained as (\ref{followerstartegyspascecase1}).
\section{Proof of Proposition 1}
1) At the RSE, the first order optimality condition holds for the follower's optimization problem as,
 \begin{equation}\label{proofoflemma1}
    \nabla_{\textbf{a}^{*RSE1}_1} u_1(\textbf{a}^{*RSE1}_1, \textbf{f}^{*RSE1}_1)=0,
 \end{equation}
where $\textbf{f}^{*RSE1}_1$ is derived by $\textbf{a}^{*RSE1}_0$. The derivative of (\ref{proofoflemma1}) with respect to $\varepsilon_1$ is
\begin{equation}\label{proofoflemma2}
     [\nabla^2_{\textbf{a}^{*RSE1}_1} u_1(\textbf{a}^{*RSE1}_1, \textbf{f}^{*RSE1}_1) \nabla_{ \varepsilon_1} \textbf{a}^{*RSE1}_1+ \nabla_{\textbf{a}^{*RSE1}_1 \textbf{f}^{*RSE1}_1}  u_1(\textbf{a}^{*RSE1}_1, \textbf{f}^{*RSE1}_1) \nabla_{\varepsilon_1} \boldsymbol{ f}_1^{*RSE1}]_{\varepsilon_1=0} =0.
 \end{equation}
If $\varepsilon_1=0$, the $u_1(\textbf{a}^{*RSE1}_1, \textbf{f}^{*RSE}_1)$ is equal to $v_1(\textbf{a}^{*NSE}_1, \textbf{f}^{*NSE}_1)$, where $\textbf{f}^{*NSE}_1$ is obtained by $\textbf{a}^{*NSE}_0$. Also, from (\ref{followerstartegyspascecase1}), the last term of left hand side of (\ref{proofoflemma2}) is equal to $-\vartheta_1^{{\scriptsize{\textnormal{T}}}}$. Therefor, by rearrangement of (\ref{proofoflemma2}), we have
\begin{equation}\label{Vraiotionof strategy}
      \nabla_{\varepsilon_1}\textbf{a}^{*RSE1}_1= (\textbf{J}^1_{\textbf{a}_1 \textbf{a}_1})^{-1} \textbf{J}^1_{\textbf{a}_1 \textbf{f}_1} \times \vartheta_{1}^{{\scriptsize{\textnormal{T}}}}.
\end{equation}
From A1, A2, and A3, the right hand side of (\ref{Vraiotionof strategy}) is negative, hence $\nabla_{\varepsilon_1}\textbf{a}^{*RSE1}_1<0$. Therefore, the follower's action is the decreasing function with respect to $\varepsilon_1$. 2) In the RSE case 1, the first order condition holds for utility of leader with respect to $\textbf{a}^{*RSE1}_1$ as
 \begin{equation}\label{proofoflemma3}
  \nabla_{\textbf{a}^{*RSE1}_0} u_0(\textbf{a}^{*RSE1}_0, \textbf{f}^{*RSE1}_0)=0.
 \end{equation}
The derivative of (\ref{proofoflemma3}) according to $\varepsilon_1$ is
\begin{eqnarray}\label{proofoflemma4}
   && [\nabla_{\textbf{a}^{*RSE1}_0}^2  u_0(\textbf{a}^{*RSE1}_0, \textbf{f}^{*RSE1}_0)\times \nabla_{\varepsilon_1}\textbf{a}^{*RSE1}_0+ \nonumber\\ && \nabla_{\textbf{a}^{*RSE1}_0 \textbf{f}^{*RSE1}_0} u_0(\textbf{a}^{*RSE1}_0, \textbf{f}^{*RSE1}_0)\times \textbf{X}_{01} \times \nabla_{\varepsilon_1} \textbf{a}_1^{*RSE1}]_{ \varepsilon_1=0} =0,
\end{eqnarray}
which is equivalent to
\begin{equation}\label{Corollary 11}
     \nabla_{\varepsilon_1}\textbf{a}^{*RSE1}_0 = -(\textbf{J}^0_{\textbf{a}_0,\textbf{a}_0})^{-1}\textbf{J}^0_{\textbf{a}_0 ,\textbf{f}_0}\boldsymbol{X}_{01}\nabla_{\varepsilon_1} \textbf{a}_{1}^{*RSE1}.
\end{equation}
From A1- A3, the right hand side of (\ref{Corollary 11}) is positive. Therefore, we have $\nabla_{\varepsilon_1}\textbf{a}^{*RSE1}_0 >0$. Hence, the leaders' action is an increasing function with respect to $\varepsilon_1$.
\section{Proof of proposition 2}
1) The Taylor series of the leader's utility around the uncertain parameter is
\begin{eqnarray}\label{tailorv0}
\lefteqn{u_0(\textbf{a}_0^{*RSE1}, \textbf{f}^{*RSE}_0)=  v_0(\textbf{a}_0^{*NSE}, \textbf{f}^{*NSE}_0)+} \\ &&\!\!\!\!\!\!\!\!\!\!\!\!\! \varepsilon_1 [( \textbf{X}_{01} \nabla_{\textbf{f}^{*RSE1}_0} u_0(\textbf{a}_0^{*RSE1}, \textbf{f}^{*RSE1}_0))^{{\scriptsize{\textnormal{T}}}} \nabla_{\varepsilon_1} \textbf{a}_{1}^{*RSE1}+(\nabla_{\textbf{a}^{*RSE1}_0} u_0(\textbf{a}_0^{*RSE1},\textbf{f}^{*RSE1}_0)))^{{\scriptsize{\textnormal{T}}}}\nabla_{\varepsilon_1} \textbf{a}_0^{*RSE1}]_{\varepsilon_1=0} +o, \nonumber
 \end{eqnarray}
Note that in the following proves, we consider the first term of Tailor series and ignore higher terms because 1) $\textbf{f}_n$ is a linear function of $\textbf{a}_{-n}$ and its higher order derivatives are equal to zero; 2) $u_n$ is a concave function of $\textbf{a}_n$ and  the second order derivative is negative, 3) the value of higher order terms is very small values because of higher exponents of $\varepsilon_1$. From A2 and $ \nabla_{\varepsilon_1}\textbf{a}_{1}^{*RSE1}<0$, the first part of (\ref{tailorv0}) is always positive. Also, the second term has only positive elements. Therefore, the utility of leader is always larger than to that of NSE in the case 1. By some rearrangements, we have
\begin{equation}\label{Corollary 12}
     \omega_{0}^{*RSE1}-\omega_{0}^{*NSE} \approx \varepsilon_1((\textbf{J}^0_{\textbf{a}_0})^{{\scriptsize{\textnormal{T}}}} \nabla_{\varepsilon_1} \textbf{a}^{*RSE1}_0+ (\textbf{X}_{10}\textbf{J}^0_{\textbf{f}_0})^{{\scriptsize{\textnormal{T}}}}\nabla_{\varepsilon_1} \textbf{a}^{*RSE1}_1).
     \end{equation}
2) Again, we start with the Taylor series of the follower's utility around $\varepsilon_1$ as
\begin{eqnarray}\label{tailorv1}
 \lefteqn{  u_1(\textbf{a}_1^{*RSE1}, \textbf{f}^{*RSE1}_1)=  v_1(\textbf{a}_1^{*NSE}, \textbf{f}^{*NSE}_1)+} \\ && \small \!\!\!\!\!\!\!\!\!\!\!\!\! \varepsilon_1 [(\textbf{X}_{01} \nabla_{\textbf{f}^{*RSE1}_1}u_1(\textbf{a}_{1}^{*RSE1}, \textbf{f}^{*RSE1}_1))^{{\scriptsize{\textnormal{T}}}}\nabla_{ \varepsilon_1} \textbf{a}_{0}^{*RSE1}+(\nabla_{\textbf{a}^{*RSE1}_{1}} u_1(\textbf{a}_{1}^{*RSE1},\textbf{f}^{*RSE1}_{1}))^{{\scriptsize{\textnormal{T}}}} \times \nabla_{\varepsilon_1} \textbf{a}_1^{*RSE1}]_{\varepsilon_1=0} +o, \nonumber
 \end{eqnarray}
In this case, the first term of the Taylor series is always negative, since $\nabla_{\textbf{f}_1} v_1(\textbf{a}_1, \textbf{f}_1)<0$ and $\nabla_{\varepsilon_1} \textbf{a}_{0}^{*RSE1}>0$. Also, $\nabla_{\textbf{a}_1} v_1(\textbf{a}_1, \textbf{f}_1)>0$ and $\nabla_{\varepsilon_1} \textbf{a}_{1}^{*RSE1}<0$. Consequently, the second term is negative. Hence, the follower's utility at RSE for case 1 is always less than the utility at NSE. By some rearrangement, we have
\begin{eqnarray}\label{Corollary 13}
  \omega_1^{*RSE1}- \omega_1^{*NSE} \approx \varepsilon_1 \times((\textbf{X}_{01}\textbf{J}^1_{\textbf{f}_1})^{{\scriptsize{\textnormal{T}}}} \nabla_{\varepsilon_1} \textbf{a}_{0}^{*RSE1}+(\textbf{J}^1_{\textbf{a}_1})^{{\scriptsize{\textnormal{T}}}}\nabla_{\varepsilon_1} \textbf{a}_{1}^{*RSE1}).
  \end{eqnarray}
3) The social utility of RSE for case 1 increases if
\begin{equation}\label{item5}
    \omega_0^{RSE1}- \omega_0^{*NSE}+ \omega_1^{*RSE1}- \omega_1^{*NSE}>0.
\end{equation}
Because $\nabla_{\varepsilon_1} \textbf{a}_{1}^{*RSE1}<0$ and $\nabla_{\varepsilon_1} \textbf{a}_{0}^{*RSE1}>0$, (\ref{item5}) leads to
\begin{eqnarray}
  \textbf{J}^0_{\textbf{a}_0}+\textbf{X}_{01}\textbf{J}_{\textbf{f}_1}^1  >0 ,\quad  \quad
 \textbf{J}^1_{\textbf{a}_1}+\textbf{X}_{10}\textbf{J}_{\textbf{f}_0}^0 <0,
\end{eqnarray}
The above conditions are equal to $\text{C}1$ and $\text{C}2$.
\section{Proof of Proposition 3}
From incomplete information set, the leader cannot calculate the exact value of $\textbf{a}_1$. Therefore, the value of $\textbf{f}_{0}$ is uncertain. Consequently, the RSE of case 2 can be considered as $\epsilon$ Stackelberg strategy space for case 1 (definition 4.7 of \cite{basar}). Since the leader's utility is continuous ( from A1 and A2), there always exists a positive value i.e. $\varsigma>0$, where $ \omega^{*RSE1}_{0}-  \omega^{*RSE2}_{0} \leq \varsigma$  (Property 4.2 of \cite{basar}) and this difference is concinnous (Property 4.3 of \cite{basar}).
\section{ Proof of Proposition 4}
1) To obtain $\omega_0^{RSE2}-\omega_0^{NSE}$, we first derive the $\nabla_{\textbf{f}^{*RSE2}_1} \textbf{a}^{*RSE2}_1$. From the first order optimality condition as $\textbf{J}^1_{\textbf{a}_1}=0$ at $\textbf{a}_1^{*RSE2}$, we have
\begin{equation}\label{lemma31}
    \textbf{J}^1_{\textbf{a}_1\textbf{a}_1} \nabla_{\textbf{f}^{*RSE2}_1} \textbf{a}^{*RSE2}_1+ \textbf{J}^1_{\textbf{a}_1\textbf{f}_1}=0,
\end{equation}
Consequently, we have $\nabla_{\textbf{f}^{*RSE2}_1} \textbf{a}^{*RSE2}_1=-(\textbf{J}^1_{\textbf{a}_1\textbf{a}_1})^{-1} \textbf{J}^1_{\textbf{a}_1\textbf{f}_1}$ which is always negative. $\nabla_{\delta_{10}} \textbf{a}^{*RSE2}_1$ is equal to
\begin{equation}\label{lemma311}
   \nabla_{\delta_{10}} \textbf{a}_1^{*RSE2}=\nabla_{\textbf{f}^{*RSE2}_1}\textbf{a}^{*RSE2}_1 \nabla_{\delta_{10}} \textbf{f}^{*RSE2}_1,
\end{equation}
From $\text{A5}$, (\ref{lemma311}) is always positive. Therefore, the follower's action is an increasing function of $\delta_{10}$. The Taylor series of $u_1(\textbf{a}^{*RSE2},\textbf{f}_1^{*RSE2})$ around $\delta_{10}$ is
\begin{equation}\label{Lemma32-a}
    u_1(\textbf{a}^{*RSE2},\textbf{f}_1^{*RSE2})=v_1(\textbf{a}^{*NSE},\textbf{f}^{*NSE}_1)+ \delta_{10} \times [\frac{\partial u_1(\textbf{a}_1^{*RSE2},\textbf{f}^{*RSE2}_1)}{\partial \delta_{10}}]_{\delta_{10}=0 , \varepsilon_1=0}+o,
\end{equation}
where
\begin{eqnarray}\label{Lemma33}
   \lefteqn{ \frac{\partial u_1(\textbf{a}_1^{*RSE2},\textbf{f}_1^{*RSE2})}{\partial \delta_{10}} =} \nonumber\\ && (\nabla_{\textbf{a}^{*RSE2}_1} u_1(\textbf{a}_1^{*RSE2},\textbf{f}^{*RSE2}_1))^{{\scriptsize{\textnormal{T}}}} \nabla_{ \textbf{f}^{*RSE2}_1}\textbf{a}^{*RSE2}_1 \nabla_{\delta_{10}} \boldsymbol{ f}^{*RSE2}_1 + \nonumber\\ &&(\nabla_{\textbf{f}^{*RSE2}_1} u(\textbf{a}_1^{*RSE2},\textbf{f}^{*RSE2}_1))^{{\scriptsize{\textnormal{T}}}} \nabla_{\delta_{10}} \textbf{f}^{*RSE2}_1,
\end{eqnarray}
and it is simplified to
\begin{equation}\label{Lemma32}
 \omega_1^{RSE2}-\omega_1^{NSE} \approx  ((\textbf{J}^1_{\textbf{a}_1})^{{\scriptsize{\textnormal{T}}}} \textbf{J}^1_{\textbf{a}_1, \textbf{f}_1} (\textbf{J}^1_{\textbf{a}_1, \textbf{a}_1})^{-1} - (\textbf{J}^1_{ \textbf{f}_1})^{{\scriptsize{\textnormal{T}}}})\nabla_{\delta_{10}} \textbf{f}^{*RSE2}_1,
\end{equation}
from A1 - A5, the follower's  utility increases with respect to $\delta_{10}$ compared to that the NSE. 2) We start with $\textbf{J}^0_{\textbf{a}_1}=0$ to derive the $\nabla_{\textbf{a}^{*RSE2}_1}\textbf{a}^{*RSE2}_0$ as,
\begin{equation}\label{lemma34}
    \textbf{J}^0_{\textbf{a}_0\textbf{a}_0}  \nabla_{\textbf{a}^{*RSE2}_1}\textbf{a}^{*RSE2}_0+ \textbf{X}_{01} \textbf{J}^0_{\textbf{a}_0\textbf{f}_0} =0.
\end{equation}
Also, the Taylor series around the uncertain information is
\begin{equation}\label{Lemma35}
    v_0(\textbf{a}_0^{*RSE2},\textbf{f}_0^{*RSE2})=v_0(\textbf{a}_0^{*NSE},\textbf{f}^{*RSE2}_0)+\delta_{10} \times [\frac{\partial v_0(\textbf{a}_0^{*RSE2},\textbf{f}_0^{*RSE2})}{\partial \delta_{10}}]_{\delta_{10}=0}+o,
\end{equation}
where
\begin{eqnarray}\label{Lemma36}
   \lefteqn{ \frac{\partial v_0(\textbf{a}_0^{*RSE2},\textbf{f}_0^{*RSE2})}{\partial \delta_{10}}=} \nonumber \\ && (\nabla_{\textbf{a}^{*RSE2}_0} v_0(\textbf{a}_0^{*RSE2},\textbf{f}^{*RSE2}_0))^{{\scriptsize{\textnormal{T}}}} \nabla_{ \textbf{a}^{*RSE2}_1} \textbf{a}^{*RSE2}_0 \nabla_{\delta_{10}} \textbf{a}^{*RSE2}_1+\nonumber\\&& (\nabla_{\textbf{f}_0} v_0(\textbf{a}_0^{*RSE2},\textbf{f}_0))^{{\scriptsize{\textnormal{T}}}}  \textbf{X}_{01} \nabla_{\delta_{10}}\textbf{a}_1+o,
\end{eqnarray}
By inserting (\ref{lemma34}) and (\ref{lemma311}) in (\ref{Lemma36}), we have
\begin{equation}\label{31}
\omega_0^{RSE2}-\omega_0^{NSE} \approx \delta_{10} \times  [-(\textbf{J}^0_{\textbf{a}_0})^{{\scriptsize{\textnormal{T}}}} \textbf{J}^0_{\textbf{a}_0, \textbf{f}_0} \textbf{X}_{01} (\textbf{J}^0_{\textbf{a}_0,\textbf{a}_0})^{-1}+ (\textbf{J}^0_{\textbf{f}_0})^{{\scriptsize{\textnormal{T}}}} \textbf{X}_{01}] \nabla_{ \delta_{10}} \textbf{a}_1 ,
\end{equation}
which is always negative from A1-A5. Now we want to derive the conditions for increasing the social utility. Since $\nabla_{\textbf{a}_1^{*RSE2}} \textbf{a}_0^{*RSE2} \times \nabla_{\delta_{10}} \textbf{a}^{*RSE2}_0 <0$, to increase the social utility, the sum of second terms in right hand side of (\ref{Lemma33}) and the first term in right hand side of (\ref{Lemma36}) should be negative as
\begin{equation}\label{first condition}
    |\textbf{J}^0_{\textbf{a}_0}| - |\textbf{J}^1_{\textbf{f}_1}| \times |\textbf{X}_{10}|<0,
\end{equation}
Since $\nabla_{\delta_{10}} \textbf{a}^{*RSE2}_1>0$, the sum of first term in right hand side of (\ref{Lemma33}) and the second term in right hand side of (\ref{Lemma36}) should be positive to increase the social utility as
\begin{equation}\label{secondcondtion}
    |\textbf{J}^1_{\textbf{a}_1}| - |\textbf{J}^0_{\textbf{f}_0}| \times |\textbf{X}_{01}|>0.
\end{equation}
Clearly, (\ref{secondcondtion}) and (\ref{first condition}) lead to $\text{C}3$ and $\text{C}4$.
\section{Proof of proposition 5}
\textbf{Lemma 2:} If $\boldsymbol{\Upsilon}$ is a P matrix, the followers' strategies are decreasing functions with respect to $\boldsymbol{\varepsilon}=[\varepsilon_1,\cdots,\varepsilon_{N_f}]$.
 \begin{proof}Assume that the followers' strategies i.e., $\textbf{a}_{\mathcal{N}_f}=[\textbf{a}_1,\cdots,\textbf{a}_{N_f}]$, are increasing functions of $\boldsymbol{\varepsilon}$, i.e.,
\begin{equation}\label{assumtionproof5}
    \textbf{a}_{\mathcal{N}_f}^{RSE1} \geq \textbf{a}_{\mathcal{N}_f}^{NSE},
\end{equation}
If $\boldsymbol{\Upsilon}$ is a $P$-matrix, $\boldsymbol{\mathcal{J}}(\textbf{a}_{\mathcal{N}_f})=(\textbf{J}^{n}_{\textbf{a}_n}(\textbf{a}_{n}))_{n=1}^{N_f}$ is strictly monotone ( Corollary 2.6.4 in \cite{palomar2010}). The equilibrium of the followers' game is a solution of $VI(\prod_{n=1}^{N_f}\mathcal{A}_{n}\times \textbf{R}_n , \boldsymbol{\mathcal{J}} )$ \cite{saeedeh5}, then,
\begin{eqnarray}
(\textbf{a}_{\mathcal{N}_f}^{RSE1} - \textbf{a}_{\mathcal{N}_f}^{NSE}) \boldsymbol{\mathcal{J}}(\textbf{a}^{NSE}_{\mathcal{N}_f})\leq 0 \label{lema222a} \\
(\textbf{a}_{\mathcal{N}_f}^{NSE} - \textbf{a}_{\mathcal{N}_f}^{RSE}) \boldsymbol{\mathcal{J}}(\textbf{a}^{NSE}_{\mathcal{N}_f})\leq 0 \label{lema222b}
\end{eqnarray}
by substituting (\ref{lema222a}) from (\ref{lema222b}) and considering (\ref{assumtionproof5}), we have
\begin{equation}\label{firstassumptions}
    \boldsymbol{\mathcal{J}}(\textbf{a}^{RSE1}_{\mathcal{N}_f}) >  \boldsymbol{\mathcal{J}}(\textbf{a}^{NSE}_{\mathcal{N}_f}),
\end{equation}
From (\ref{followerstartegyspace}),
\begin{equation}\label{jacobielements}
   \frac{\partial u_n^k(a_n^k, f_n^k)}{\partial a_n^k}= \frac{\partial v_n^k(a_n^k, \widetilde{f}_n^{k*})}{\partial a_n^k} + \frac{\partial v_n^k}{\partial\widetilde{f}_n^{k*}} \times \frac{\partial \widetilde{f}_n^{k*}}{\partial a_n^k},
\end{equation}
and
 \begin{equation}\label{100}
   \frac{\partial \widetilde{f}_n^{k*}}{\partial a_n^k}=  \frac{\partial \widetilde{f}_n^{k*}}{ \partial \vartheta^k_n} \times \frac{\partial \vartheta^k_n}{\partial a_n^k } =  -\varepsilon_1 \times \frac{\partial^2 v^k_n(\textbf{a}_n, \widetilde{f}^*_1)}{\partial a_n^k \widetilde{f}_1^k}  \times (\sum_{k=1}^{K} (\frac{\partial u^k_1(\textbf{a}_1, \widetilde{\textbf{f}}_1)}{\partial f_1^k})^2)^{-\frac{1}{2}},
 \end{equation}
From $ \widetilde{a}_n^k = -\varepsilon_1 \times \frac{\partial v_n^k}{\partial\widetilde{f}_n^{k*}} \times \frac{\partial^2 v^k_n(\textbf{a}_n, \widetilde{f}^*_1)}{\partial a_n^k \widetilde{f}_1^k}  \times (\sum_{k=1}^{K} (\frac{\partial u^k_1(\textbf{a}_1, \widetilde{\textbf{f}}_1)}{\partial f_1^k})^2)^{-\frac{1}{2}}$ which is negative according to A1-A3, (\ref{100}) is equal to
\begin{equation}\label{}
    \boldsymbol{\mathcal{J}}(\textbf{a}^{RSE1}_{\mathcal{N}_f}) - \boldsymbol{\mathcal{J}}(\textbf{a}^{NSE}_{\mathcal{N}_f}) =\widetilde{\textbf{a}} <\textbf{0},
\end{equation}
where $\widetilde{\textbf{a}}=(\widetilde{\textbf{a}}_n)_{n=1}^{N_f}$, $\widetilde{\textbf{a}}_n^{\text{T}}=[\widetilde{a}_n^1,\cdots,\widetilde{a}_n^K]$ which is contradict with (\ref{firstassumptions}). This contradiction implies that our assumption is wrong. Therefore, the followers' actions in the Robust Stackelberg game case 1 are decreasing functions with respect to $\boldsymbol{\varepsilon}$. \end{proof}
1) Since the followers' strategies are decreasing functions of $\boldsymbol{\varepsilon}$, the value of $\textbf{f}_0$ decreases which implies $v_0^{RSE1} \geq v_0^{NSE}$ from A2. Consider the variation of the leader's action with respect to the bound of uncertainty region of all followers as $\nabla_{\boldsymbol{\varepsilon}} \textbf{a}_0$. The Taylor series of $v_0^{RSE1}$ around $\boldsymbol{\varepsilon}$ is
\begin{equation}\label{Lemma25}
    v_0^{RSE1}\approx v_0^{NSE} +[ (\nabla_{\boldsymbol{\varepsilon}} \textbf{a}_0)^{{\scriptsize{\textnormal{T}}}}\textbf{J}^{n}_{\textbf{a}_0}+ \sum_{n=1}^{N_f} \varepsilon_{n} \times [ \textbf{X}_{0n} \textbf{J}^{0}_{\textbf{f}_0} (\nabla_{\varepsilon_{n}}\textbf{a}_{n})^{{\scriptsize{\textnormal{T}}}} ]+o.
\end{equation}
2) The RNE of the followers in multiple-followers of RSG in Section V belongs to Robust additively coupled game introduced in \cite{saeedeh5}. Based on Theorem 2 of \cite{saeedeh5}, the social utilities of the followers at RSE is less than that at NSG, if $\boldsymbol{\Gamma}$ is P matrix. The Taylor series around $\boldsymbol{\varepsilon}$ is
\begin{equation}\label{Lemma253}
    v_{n}^{RSE1}\approx v_n^{NSE} + \varepsilon_{n} \times[ (\textbf{J}^{n}_{\textbf{f}_n})^{{\scriptsize{\textnormal{T}}}}  \textbf{X}_{n 0} \nabla_{\boldsymbol{\varepsilon}} \textbf{a}_0+ \sum^{N_f}_{m=1 , m \neq n} \textbf{X}_{n m} \nabla_{\varepsilon_{m}} \textbf{a}_m+  (\textbf{J}_{\textbf{a}_n}^n)^{{\scriptsize{\textnormal{T}}}}  \nabla_{\varepsilon_{n}}\textbf{a}_n ]+o.
\end{equation}
3) The social utility at RSE case 1 increases, if the summation of (\ref{Lemma25}) and (\ref{Lemma253}) is positive. Therefore the terms multiplied by $\nabla_{\varepsilon_n} \textbf{a}_0$ should be positive because $\nabla_{\boldsymbol{\varepsilon}} \textbf{a}_0>0$. Since $\nabla_{\varepsilon_n} \textbf{a}_n<0$, its multiplied terms should be negative. These two conditions leads to $\text{C5}$ and $\text{C6}$ by some rearrangements.
\section{Proof of proposition 6}
\textbf{Lemma 3:} If $\boldsymbol{\Upsilon}$ is a P matrix, the followers' strategies are increasing functions with respect to $\boldsymbol{\delta}_0=[\delta_{10},\cdots, \delta_{N_f,0}]$.
 \begin{proof} Assume the follower's strategies i.e., $\widetilde{\textbf{a}}_{\mathcal{N}_f}=[\textbf{a}_1,\cdots,\textbf{a}_{N_f}]$, are the decreasing functions of $\boldsymbol{\delta}_0$. Therefore, we have $\widetilde{\textbf{a}}_{\mathcal{N}_f}^{RSE2} \leq \widetilde{\textbf{a}}^{NSE}$. If $\boldsymbol{\Upsilon}$ is a P matrix, $\boldsymbol{\mathcal{J}}(\textbf{a}_{\mathcal{N}_f})=(\textbf{J}^{n}_{\textbf{a}_n}(\textbf{a}_{n}))_{n=1}^{N_f}$ is strictly monotone ( Theorem 12.5 in \cite{palomar2010}), and we have,
\begin{equation}\label{Lemma22}
   \boldsymbol{\mathcal{J}}(\textbf{a}^{RSE2}_{\mathcal{N}_f}) \leq  \boldsymbol{\mathcal{J}}(\textbf{a}^{NSE}_{\mathcal{N}_f})
\end{equation}
Now, from the Taylor series of the followers' utilities round $\boldsymbol{\delta}_0$, we have
\begin{equation}\label{jacobielements35}
   \frac{\partial u_n^k(a_n^k, f_n^k)}{\partial a_n^k}= \frac{\partial v_n^k(a_n^k, \widetilde{f}_n^{k*})}{\partial a_n^k} + \frac{\partial v_n^k}{\partial\widetilde{f}_n^{k*}} \times \frac{\partial \widetilde{f}_n^{k*}}{\partial \delta_{n0}},
\end{equation}
from A5, the last term of (\ref{jacobielements35}) is positive. Therefore
\begin{equation}\label{Lemma36-2}
   \boldsymbol{\mathcal{J}}(\textbf{a}^{RSE2}_{\mathcal{N}_f}) -  \boldsymbol{\mathcal{J}}(\textbf{a}^{NSE}_{\mathcal{N}_f})=\widetilde{\textbf{b}}>0
\end{equation}
where $\widetilde{\textbf{b}}=(\widetilde{\textbf{b}}_n)_{n=1}^{N_f}$ and $k^{\text{th}}$ elements of $\widetilde{\textbf{b}}_n$ is equal to the last term of (\ref{jacobielements35}). From (\ref{Lemma36}), $\boldsymbol{\mathcal{J}}(\textbf{a}^{RSE2}_{\mathcal{N}_f})\geq \boldsymbol{\mathcal{J}}(\textbf{a}^{NSE}_{\mathcal{N}_f})$ which is contradict with (\ref{Lemma22}). Therefore the followers' strategies are increasing functions with respect to $\boldsymbol{\delta}_{0}$.
 \end{proof}
1) If the followers' actions are increasing functions of $\boldsymbol{\delta}_0$, the value of $\textbf{f}_0$ increases. Since $v_0$ is a decreasing function of $\textbf{f}_0$, the leader's utility is a decreasing function with respect to $\boldsymbol{\delta}_0$. By using the Taylor series, we have
\begin{equation}\label{Lemma256}
    v_0^{NSE2}\approx v_0^{NSE} + \sum_{n \in \mathcal{N}_f}\delta_{n0} \times ( (\textbf{J}^0_{\textbf{f}_0})^{\text{T}} \textbf{X}_{0n} \nabla_{\delta_{n0}} \textbf{a}_n)+ (\textbf{J}^0_{\textbf{a}_0})^{\text{T}} \nabla_{\delta_{n0}}\textbf{a}_0)+o.
\end{equation}
2) Assume that the followers' utilities are decreasing functions with respect to $\boldsymbol{\delta}_0$. In this case, the followers' strategies are decreasing functions with respect to $\boldsymbol{\delta}_0$. This is because $\textbf{J}^n_{\textbf{a}}$ is strong monotone when $\boldsymbol{\Upsilon}$ is P-matrix. However, this is contradict with Lemma 3. Therefore, the followers' utilities are increasing function with respect to $\boldsymbol{\delta}_0$. Consequently, the social utility of game converges to the higher utility. Besides, the Taylor series of each follower's utility around $\delta_{n0}$ is
\begin{equation}\label{Lemma26}
    v_n^{RSE2}\approx v_n^{NSE} + \delta_{n0} \times [ (\textbf{J}_{\textbf{f}_n}^{n})^{\text{T}} \textbf{X}_{n0} \nabla_{\delta_{0n}} \textbf{a}_0 + (\textbf{J}_{\textbf{f}_n}^{n})^{\text{T}} (\sum_{m\neq n, \, m \in \mathcal{N}_f} \textbf{X}_{nm} \nabla_{\delta_{m0}} \textbf{a}_m)+ (\textbf{J}^n_{\textbf{a}_n})^{\text{T}} \nabla_{\delta_{0n}} \textbf{a}_n]+o.
\end{equation}
3) The social utility of RSE2 increases compared to that at the NSE, if the summation of second part of Taylor series in (\ref{Lemma25}) and (\ref{Lemma26}) are positive. In this case, $\nabla_{\delta_{n0}} \textbf{a}_0<0$ and $\nabla_{\delta_{n0}} \textbf{a}_n >0$, therefore the terms related to the  $\nabla_{\boldsymbol{\delta}_{0}} \textbf{a}_0$ should be negative, and the terms related to $\nabla_{\delta_{n0}} \textbf{a}_n$ should be positive. By some rearrangements, the $\text{C7}$ and $\text{C8}$ are obtained.
\section{Proof of Proposition 7}
By introducing robustness in the followers' optimization problems, their strategies at the RSE decrease [Theorem 2 in \cite{saeedeh5}]. Consequently, the followers' impacts on the leaders decrease, and the leaders' utilities increase. Therefore, the leaders' social utility increases.
\small \small
\begin{table*}[et]\label{table2}
\centering
\begin{center}
    \begin{tabular}{ | c | c | c | c |}
    \hline
    Case of robust game & Leader & Follower & The social utility increases if   \\ \hline
    case 1 & $\omega_{0}^{RSE1}\geq \omega_0^{NSE}$ &  $ \omega_1^{RSE1} \leq \omega_1^{NSE}$ & $    |\textbf{J}^{0}_{\textbf{a}_0}|>|\textbf{C}_{10}| \, ,  |\textbf{J}^{1}_{\textbf{a}_1}|< |\textbf{C}_{01}|$ \\ \hline
    case 2 & $  \omega_{0}^{RSE2} \leq \omega_{0}^{NSE}$ &  $ \omega_1^{RSE2} \geq \omega_1^{NSE}$ & $ |\textbf{J}^{0}_{\textbf{a}_0}| < | \textbf{C}_{10}| ,\, | \textbf{J}^{1}_{\textbf{a}_1}| > | \textbf{C}_{01}|$ \\ \hline
         \end{tabular}
\end{center}
\caption{Summary of Propositions 2 and 4.}\label{tablesummery}
\end{table*}
\small \small
\begin{table*}[gt]
\centering
\begin{tabular}{| c | c | c | c | c| c|c| c|}
  \hline
  Number of Fig & $N_l$ & $N_f$ & $K$ & $H_{mn}^{k}$ & $a_{nk}^{\text{max}}$ &  $a_{nk}^{\text{min}}$ & $\sigma^k_{n}$ for all $k$\\ \hline
  \ref{fig5} & $1$& $1$  &$1$  &  A Rayleigh model \cite{newprespetive}  & $10$ dB & $0$ & $0.01$\\
 \ref{Figcase1}-\ref{Figcase2} & $1$ &  $1$&  $20$ & A four-ray Rayleigh model \cite{newprespetive} & $10$ dB & $0$ & $0.01$ \\
  \hline
\end{tabular}
\caption{Simulation parameters for power control games}\label{simulation}
\end{table*}
\small \small
\begin{table*}[ht]
\centering
\begin{tabular}{|c||c|c|c||c|c|c||c|c|c|}
\hline
Different scenarios based on SINR &\multicolumn{3}{c|}{R1}& \multicolumn{3}{c|}{R2}&\multicolumn{3}{c|}{R3}\\
\cline{2-10}\hline
 achieved utility for case 1 &RSE&NSE& $d\%$&RSE&NSE&$d\%$&RSE&NSE&$d\%$\\
\hline
leader & $92.17$& $71.95$& $28.1\%$& $110.11$&  $105.71$   &$4.2\%$& $156.01$ & $119.89$ & $30.13\%$ \\
Follower & $53.47$& $70.84$&$-24.5\%$& $97.67$ & $95.94$ &$1.8\%$&$50.15$&$96.55$ & $-48.06\%$\\
\hline
\end{tabular}
\caption{The leader's and the follower's utility subject to (\ref{powerlimitation}).}\label{Tableutilitypower}
 \end{table*}
\small \small
 \begin{table}[ht]
\centering
\begin{tabular}{l}
\hline\hline
\textbf{Heuristic protocol for multiple-leaders and multiple-followers}\\
\hline \hline
\textbf{Start $n_l=1$ and $n_l \in N_l$},
\\Consider $\mathcal{N}_{n_f}^{new}=\{1,...,n_{l-1},n_{l+1},...,N_l\}\bigcup \mathcal{N}_f$,
\\ For all the players of Stackelberg game except leader $n_l$,
\\ Calculate $\text{C7}$-$\text{C8}$ by considering $n_l=0$ and $\mathcal{N}_{n_f}^{new}$ as the set of followers,
\\ If $\text{C7}$-$\text{C8}$ hold for $n_l$, \\ Play RSG with $n_l$ as leader and others as followers:
\\ 1) Leader $n_l$ announces its strategy,
   \\ 2) All the players of $\mathcal{N}_{n_l}$ play the strategic game,
   \\3) break.
\\otherwise $n_l=n_l+1$,
\\continue ,
\\ If $n_l=N_l+1$,\\ End.
\\ \hline
\end{tabular}\caption{Heuristic protocol for case 2 of multiple-leaders and multiple-followers.  }  \label{table4}
\end{table}
\begin{table}[h]
\centering
\begin{center}
    \begin{tabular}{ | c | c | c | c |}
    \hline
    Utility  & NE & NSE & RSE for heuristic algorithm  \\ \hline \hline
    Leader 1 & $7.07$ & $5.06$ & $4.97$    \\ \hline
    Leader 2 & $ 1.67$ &  $ 4.33$ & $4.19$  \\ \hline
    Follower & $1.7$ &  $2.08 $ & $ 2.49$  \\ \hline
       Social utility of the leader 2 and the follower &$ 3.37$ & $6.41$& $6.68$ \\ \hline \end{tabular}
\end{center}
\caption{Performance of the heuristic protocol.}\label{tablehuresticalgorrithm}
\end{table}
\end{document}